\documentclass[%
 reprint,
 superscriptaddress,preprintnumbers,
 nofootinbib,
 amssymb,
 aps,
]{revtex4-2}

\usepackage{amsmath}
\usepackage{dsfont}
\usepackage[utf8]{inputenc}
\usepackage{hyperref}
\usepackage{graphicx}   
\usepackage[table]{xcolor}
\usepackage[dvipsnames]{xcolor}
\usepackage{colortbl}
\usepackage{url}
\usepackage[normalem]{ulem}
\usepackage{slashed}
\usepackage[capitalise, english]{cleveref}
\usepackage[noindentafter]{titlesec} 
\usepackage{bm}
\usepackage{tikz}
\usepackage[compat=1.1.0]{tikz-feynman}
\usepackage[compat=1.1.0]{tikz-feynhand}

\usepackage{booktabs}
\usepackage{graphicx}
\usepackage{amssymb}
\usepackage{xcolor}
\usepackage{pifont}
\usepackage{fontawesome}

\usepackage{booktabs}
\usepackage{siunitx}
\usepackage{xspace}
\usepackage{enumitem}

    \makeatletter
    \def\CT@@do@color{%
      \global\let\CT@do@color\relax
            \@tempdima\wd\z@
            \advance\@tempdima\@tempdimb
            \advance\@tempdima\@tempdimc
    \advance\@tempdimb\tabcolsep
    \advance\@tempdimc\tabcolsep
    \advance\@tempdima2\tabcolsep
            \kern-\@tempdimb
            \leaders\vrule
                    \hskip\@tempdima\@plus  1fill
            \kern-\@tempdimc
            \hskip-\wd\z@ \@plus -1fill }
    \makeatother

\usepackage{accents}
\DeclareMathSymbol{\widetildesym}{\mathord}{largesymbols}{"65}

\def\thesubsection{\arabic{section}.\arabic{subsection}}

\def\thesection{\arabic{section}}
\titleformat*{\subsubsection}{\normalfont \small \bfseries \boldmath}
\renewcommand{\paragraph}[1]{\vspace{.3em} \indent {\bfseries \boldmath #1 ---}\xspace }
\makeatletter
    \renewcommand{\p@subsection}{}
    \renewcommand{\p@subsubsection}{}
\makeatother

\definecolor{red}{rgb}{0.6,.0706,.1373}
\definecolor{blue}{rgb}{0,0.396,0.741}
\newcommand\myshade{80}
\colorlet{mylinkcolor}{violet}
\colorlet{mycitecolor}{violet}
\colorlet{myurlcolor}{violet}

\hypersetup{
  linkcolor  = mylinkcolor!\myshade!black,
  citecolor  = mycitecolor!\myshade!black,
  urlcolor   = myurlcolor!\myshade!black,
  colorlinks = true
}
\newcommand{\eminus}{\vcenter{\hbox{\scalebox{0.6}[1]{$ - $}}}}	

\newcommand{\sscript}[1]{{\scriptscriptstyle \mathrm{#1}}}

\newcommand{\be}{\begin{eqnarray}}
\newcommand{\ee}{\end{eqnarray}}
\newcommand{\beq}{\begin{equation}}
\newcommand{\eeq}{\end{equation}}

\newcommand{\sig}{\sigma}
\newcommand{\lzm}{\left(}
\newcommand{\dzm}{\right)}

\newcommand{\cL}{\mathcal{L}}
\newcommand{\cO}{\mathcal{O}}

\newcommand{\cC}{{\mathcal C}}
\newcommand{\cS}{{\mathcal S}}
\newcommand{\cP}{{\mathcal P}}

\newcommand{\mev}{\mathrm{MeV}}
\newcommand{\gev}{\mathrm{GeV}}
\newcommand{\tev}{\mathrm{TeV}}

\newcommand{\hermc}{\text{h.c.}}

\newcommand{\cmark}{{\color{black}\ding{51}}}

\keywords{}

\begin{document}

\title{
\boldmath Expanding the Landscape of Exotic Muon Decays
}

\author{Admir Greljo}
\email{admir.greljo@unibas.ch}
\affiliation{Department of Physics, University of Basel, Klingelbergstrasse 82,  CH-4056 Basel, 
Switzerland}

\author{Ajdin Palavri\'c}
\email{ajdin.palavric@unibas.ch}
\affiliation{Department of Physics, University of Basel, Klingelbergstrasse 82,  CH-4056 Basel, 
Switzerland}

\author{Mirsad Tunja}
\email{mirsad.tunja@pmf.unsa.ba}
\affiliation{Faculty of Science, University of Sarajevo, Zmaja od Bosne 33-35, 71000 Sarajevo, Bosnia and Herzegovina}

\author{Jure Zupan}
\email{zupanje@ucmail.uc.edu}
\affiliation{Department of Physics, University of Cincinnati, Cincinnati, Ohio 45221, USA}


\begin{abstract}

We chart new-physics models that produce exotic, high-multiplicity muon decays featuring prompt or displaced $e^+e^-$ pairs and/or photons, with or without missing energy, such as $\mu \to 5e$, $\mu \to 7e$, etc. Starting from an effective-field-theory perspective, we estimate the reach on the ultraviolet scale and identify conditions under which lower-multiplicity modes are suppressed or occur at comparable rates. We then construct explicit realizations in minimal dark-sector models with light, feebly interacting particles, such as flavor-protected scalars, dark photons, inelastic dark matter, and axion-like particles. The predicted novel signatures can be probed at MEG~II and Mu3e, as well as during calibration runs of COMET and Mu2e. A future discovery would provide valuable insights into short-distance dynamics and the mechanism of lepton-flavor symmetry breaking.

\end{abstract}

\maketitle

\tableofcontents

\section{Introduction} 
\label{sec:intro}

In the Standard Model (SM) of particle physics, individual lepton flavors are conserved in tau and muon decays to an excellent approximation, broken only by corrections proportional to the tiny neutrino masses. Any observation of charged-lepton flavor violation (cLFV) would therefore provide a smoking gun signal of New Physics (NP). The three golden rare muon transitions of this type are $\mu \to e \gamma$ and $\mu \to eee$ decays~\cite{Feinberg:1958zzb}, and muon-to-electron conversion on nuclei, $\mu^- A \to e^- A^{(*)}$~\cite{Marciano:1977cj}. There is a concerted effort to significantly improve the experimental searches for these transitions~\cite{COMET:2025sdw, Corrodi:2023mza}. By the end of the decade, the Mu2e experiment at Fermilab~\cite{Mu2e:2012eea, Mu2e:2014fns, Mu2e-II:2022blh, CGroup:2022tli} and the COMET experiment at J-PARC~\cite{Kuno:2013mha, COMET:2018auw} are expected to be sensitive to $\mu \to e$ conversion rates that are over four orders of magnitude below the current bounds. In contrast, the DeeMe experiment at J-PARC~\cite{Teshima:2019orf} is expected to improve bounds by an order of magnitude on a shorter timescale. The MEG-II experiment at PSI is currently taking and analyzing data, steadily tightening the bounds on $\mu \to e\gamma$~\cite{MEGII:2023ltw, MEGII:2025gzr}. It is projected to improve the sensitivity by nearly an order of magnitude compared to its predecessor. Likewise, the Mu3e experiment, also at PSI, is expected to improve the current limits on the $\mu \to 3e$ decay rate by about four orders of magnitude by the end of its Phase-II~\cite{Blondel:2013ia}. This experimental program is expected to be one of the main pillars of the field in the near future~\cite{deBlas:2944678}.

\begin{table}[t]
\begin{center}
\renewcommand{\arraystretch}{1.7}  
\setlength{\tabcolsep}{12pt}       
\begin{tabular}{c | c c c c c}
$\bm{e \backslash \gamma}$ & \textbf 0 & \textbf 1 & \textbf 2 & \textbf 3 & \textbf 4 \\ \hline
\textbf 1 & $e$ & $e\gamma$ & \textcolor{red}{$e2\gamma$} & \textcolor{blue}{$e3\gamma$}  & \textcolor{green!70!black}{$e4\gamma$} \\
\textbf 3 & $3e$ & \textcolor{red}{$3e\gamma$} & \textcolor{blue}{$3e2\gamma$} & \textcolor{green!70!black}{$3e3\gamma$} &\dots \\
\textbf 5 & \textcolor{red}{$5e$} & \textcolor{blue}{$5e\gamma$} & \textcolor{green!70!black}{$5e2\gamma$} &\dots  & \\
\textbf 7 & \textcolor{blue}{$7e$} & \textcolor{green!70!black}{$7e\gamma$} &\dots & & \\
\textbf 9 & \textcolor{green!70!black}{$9e$} &\dots & & &
\end{tabular}
\caption{An overview of exotic muon decays organized by the number of electrons and positrons (rows) versus the number of photons (columns) in the final state. The well-known transitions, $\mu \to e$ conversion, $\mu \to e \gamma$, and $\mu \to 3e$ are denoted in black. The modes that contain a single extra building block ($e^+e^-$ pair or a photon) are denoted in red; those with two (three) extra building blocks are in blue (green). Each entry may additionally include missing energy, while tracks may be prompt or displaced. }
\label{tab:exotic_signatures}
\end{center}
\end{table}

In terms of sensitivity to short-distance physics beyond the SM captured by the dimension-6 SM effective field theory (SMEFT) framework, these experiments will be able to probe scales up to $\Lambda \sim 10^8$ GeV; see Refs.~\cite{Ardu:2024bua, Fernandez-Martinez:2024bxg, Calibbi:2017uvl} for a recent global analysis in this context.

If there is a light, feebly interacting particle in the spectrum, rare muon decays involving such light particles can indirectly probe even higher scales~\cite{Agrawal:2021dbo, Antel:2023hkf, Tammaro:2025zso}. A well-motivated example is a pseudo-Nambu-Goldstone boson (pNGB) $a$, arising from a spontaneously broken approximate global symmetry, which couples derivatively to the SM currents. The flavor-violating dimension-5 operator $(\partial_\mu a /f)(\bar\mu\gamma^\mu\gamma_5 e)$ induces $\mu\to ea$ decays with a parametric enhancement of the branching ratio, i.e., $\mathrm{BR}(\mu\to ea)\propto (m_W^2/m_\mu f)^2$. Even a relatively weak bound $\mathrm{BR}(\mu\to ea)\lesssim\mathcal O(10^{\eminus6})$ already implies $f\gtrsim10^{10}\,\mathrm{GeV}$, as in scenarios where $a$ decays invisibly and large SM backgrounds are present.
The sensitivity only improves in the case of more exotic $a$ decay modes, which have strongly suppressed backgrounds. Future searches with sensitivities to exotic muon decay branching ratios at the level $\mathcal O(10^{\eminus15})$ could then probe scales as high as $f\sim10^{14}\,\mathrm{GeV}$. This motivates a detailed exploration of light NP scenarios testable in rare muon decays. Examples studied in the literature include invisible axion-like particle (ALP) production in $\mu \to ea$ two-body decays~\cite{GarciaiTormo:2011cyr, Uesaka:2020okd, Calibbi:2020jvd, Panci:2022wlc, Jho:2022snj, Xing:2022rob, Hill:2023dym, Knapen:2023zgi, Bigaran:2025uzn}, virtual ALP exchange~\cite{Fuyuto:2023ogq}, and muon-decay-induced production of light particles that subsequently decay to visible SM states such as $e^\pm$ and photons~\cite{Echenard:2014lma, Heeck:2017xmg, Hostert:2023gpk, Knapen:2023iwg, Fox:2024kda, Knapen:2024fvh}, or stay invisible~\cite{Jahedi:2025hnu}. For a broader overview, see Ref.~\cite{MartinCamalich:2025srw}.

In this paper, we adopt a signature-driven approach to rare muon decays. The possible exotic signatures can be systematically classified using a finite set of building blocks: additional $e^+e^-$ pairs, extra photons, and missing energy, as summarized in \cref{tab:exotic_signatures}. If long-lived resonances appear in the decay chain, displaced vertices may also arise. Several of these signatures, particularly those involving higher-multiplicity final states, remain unexplored both theoretically and experimentally. Our goal is to assess their relevance for physics beyond the SM, provide benchmark models, and motivate dedicated searches at MEG-II and Mu3e, as well as using calibration runs at Mu2e and COMET~\cite{Hill:2023dym}.

In the SM, higher-multiplicity decays arise from QED radiation and can be efficiently suppressed in experimental searches through suitable kinematical cuts and the use of precise theoretical predictions~\cite{Colangelo:2021xix, Banerjee:2022nbr}. A key question is how complex an NP model must be for such signatures to compete with, or even dominate over, the lower-multiplicity channels. Addressing this can help guide experimental priorities and clarify which signatures could be incorporated into existing searches, potentially with only modest adjustments to the experimental setup.

The structure of this paper is as follows. In \cref{sec:EFT_considerations}, we adopt a systematic effective field theory (EFT) approach to identify the lowest-dimensional operators that can mediate exotic muon decays, both within the SMEFT and in scenarios with light NP. We estimate the effective scales probed by each class of operators, finding that scenarios with light states offer particularly promising avenues for exploration. Building on these insights, \cref{sec:models} introduces four renormalizable benchmark models that realize the exotic decay signatures.
These include processes such as $\mu \to 7e$, $\mu \to 3e (+ \mathrm{inv})$, and $\mu \to e\gamma\gamma$, which can serve as complementary probes. From a model-building perspective, these benchmarks highlight the crucial role played by the lepton flavor symmetries and their breaking patterns. On the experimental side, we discuss promising search strategies, outline optimal analysis approaches, and provide {\tt UFO} models to facilitate dedicated experimental studies. We conclude in \cref{sec:conclusions}.

\section{Effective Field Theory Considerations}
\label{sec:EFT_considerations}

Our EFT analysis is structured around two conceptually distinct scenarios. In \cref{sec:SMEFT}, we first consider the case where all new states are heavier than the electroweak scale $\langle H \rangle = v_\text{ew}/\sqrt2$, $v_\text{ew}\simeq246$\,GeV. They can thus be integrated out and the theory described by the local SMEFT operators. As a part of \cref{sec:SMEFT}, we also discuss in passing the intermediate regime characterized by $m_\mu < m_X < v_{\text{ew}}$, where $X$ is too heavy to be produced in muon decays, yet too light for a true SMEFT description to be valid.  In \cref{sec:EFT:lightNP}, we then analyze the possibility that new SM-singlet particles $X$ are sufficiently light to be produced in muon decays, $m_X < m_\mu-m_e$. In this regime, we formulate SMEFT extensions involving various representative light fields $X$ uncharged under the SM, and identify gauge-invariant operators that couple SM fields to the light state $X$.

We mainly focus on model-building aspects and side-step detailed experimental questions. Unless otherwise stated, we use $\mathrm{BR}(\mu\to e+X)<10^{\eminus 15}$ as an illustrative benchmark value for all higher-multiplicity decays, reserving dedicated experimental sensitivity projections for each signature for future studies. The quoted numerical values should thus be taken only as rough guidance, even though sensitivity to such small branching ratios may well be experimentally achievable for many exotic channels already with the next generation of experiments.

The reason is that in many cases, the searches will face only highly suppressed backgrounds. In the SM, high-multiplicity muon decays arise due to radiative corrections. That is, the Michel decay $\mu \to e \nu_\mu \bar{\nu}_e$ gets  corrected at ${\mathcal O}(\alpha/\pi)$ by an emission of a photon, giving $\mu \to e \nu_\mu \bar{\nu}_e \gamma$. The emitted photon can also be off-shell, which then Dalitz converts to an $e^+e^-$ pair, giving rise to $\mu \to 3 e \nu_\mu \bar{\nu}_e$ decays. Repeated occurrence of this splitting yields final states with multiple electrons and photons, with each addition of $\gamma$ and/or $e^+e^-$ pair in the final state resulting in further suppressed branching ratios. As an example, the SM branching ratio for $\mu \to 3e2\nu$ is $\cO(10^{\eminus5})$, while for $\mu\to 5 e 2\nu$ it is $\cO(10^{\eminus10})$ \cite{Hostert:2023gpk}. Note that the contributions to high-multiplicity muon decays due to electroweak corrections are negligible. For example, the leading ${\mathcal O}(G_F^2)$ contribution gives $\mathrm{BR}(\mu\to 3e2\nu)\sim\cO(10^{\eminus 20})$.

The SM decays $\mu \to  (m e) (n\gamma)\nu_\mu\bar \nu_e$, with $m,n$ integers ($m$ being odd), constitute irreducible backgrounds to the NP searches. The kinematics of the SM and NP decays, however, differ in general. For example, the photon energy spectra in the SM decays exhibit an infrared divergence, which is regulated by experimental constraints, such as the minimum transverse momentum threshold for the photon, $p_T^{\sscript{min}}$.
In NP decays, on the other hand, the photons can be, on average, significantly harder, with the typical energy set by, e.g., the decays of light NP particles. If the NP signal contains no missing energy, the SM background can be especially efficiently suppressed using kinematical cuts on missing momentum.

In many cases, especially when exotic channels involve missing energy, it is crucial to have precise predictions for the SM backgrounds to perform an NP search effectively. The precision computations of several relevant differential SM rates can be found in~\cite{Fael:2016yle, Banerjee:2020rww, Pruna:2016spf}. To determine the experimental sensitivities, additional reducible backgrounds, such as photon conversions in a material and coincidences between multiple muon decays, must also be considered~\cite{Mu3e:2020gyw, Echenard:2014lma, Hostert:2023gpk}. For high-multiplicity exotic muon decays, all sources of SM backgrounds decrease quickly, either because the relevant SM branching ratios decrease rapidly with increasing multiplicity or because it becomes harder to achieve coincidences when many particles are involved, quickly reaching values far below the current and near-term experimental sensitivities.

\subsection{Standard Model Effective Field Theory}
\label{sec:SMEFT}

\begin{table}[t]
\centering
\renewcommand{\arraystretch}{1.5}
\scalebox{0.9}{
\begin{tabular}{c@{\hspace{1.25cm}}c@{\hspace{1.25cm}}c}
\toprule\toprule
\textbf{Signature}
&$\bm{\cL_{\sscript{SMEFT}}\supset}$
&$\bm{\Lambda\,[\gev]}$
\\
\midrule
$\bm{\mu\to e\gamma}$
&$\frac{y_\mu}{16\pi^2}\frac{1}{\Lambda^2}(\bar\ell_2\sigma^{\mu\nu}e_1)HB_{\mu\nu}$
&$3\times10^5$
\\
$\bm{\mu\to 3e}$
&$\frac{1}{\Lambda^2}(\bar\ell_2\gamma^\mu\ell_1)(\bar\ell_1\gamma_\mu\ell_1)$
&$2\times10^6$
\\
\midrule
$\bm{{\color{red}{\mu\to 5e}}}$
&$\frac{y_\mu}{\Lambda^{6}}(\bar\ell_1\gamma^\mu\ell_1)^2(\bar\ell_2 He_1)$
&$5$
\\
$\bm{{\color{red}{\mu\to3e\gamma}}}$
&$ \frac{1}{16\pi^2}\frac{1}{\Lambda^4}(\bar\ell_2\gamma^\mu\ell_1)(\bar\ell_1\gamma^\nu\ell_1)B_{\mu\nu}$
&$30$
\\
$\bm{{\color{red}{\mu\to e2\gamma}}}$
&$\frac{y_\mu}{16\pi^2}\frac{1}{\Lambda^4}(\bar\ell_2 He_1)B_{\mu\nu}B^{\mu\nu}$
&$80$
\\
\midrule
$\bm{{\color{blue}{\mu\to e3\gamma}}}$
&$\frac{1}{16\pi^2}\frac{1}{\Lambda^4}(\bar \ell_2\gamma^\mu\overleftrightarrow D^\nu \ell_1)B_{\mu \rho} B_\nu^\rho$
&$20$
\\
$\bm{{\color{blue}{\mu\to 3e2\gamma}}}$
&$\frac{1}{\Lambda^4}(\bar \ell_2\gamma^\mu\overleftrightarrow D^\nu \ell_1)(\bar \ell_1 \gamma_\mu \overleftrightarrow D_\nu \ell_1)$
&$40$
\\
\bottomrule\bottomrule
\end{tabular}
}
\caption{Overview of the representative SMEFT operators contributing to rare muon decay channels along with the scales $\Lambda$ they probe. For the golden modes, $\mu \to e\gamma$ and $\mu \to 3e$, the quoted sensitivities are obtained using the projected bounds on their branching ratios, i.e. $\mathrm{BR}(\mu \to e\gamma)\sim6\times10^{\eminus14}$ and $\mathrm{BR}(\mu \to 3e)\sim10^{\eminus16}$~\cite{Ardu:2024bua, Calibbi:2021pyh}. For all of the remaining higher-multiplicity modes, the lower bound on the scale is extracted assuming a branching-ratio sensitivity of $\mathrm{BR}(\mu\to e+X)<10^{\eminus15}$. Normalization conventions for different operators are discussed in \cref{sec:SMEFT}.
}
\label{tab:SMEFT_bounds}
\end{table}

Let us first assume that all NP  states are heavy, so that they can be integrated out and the theory matched onto SMEFT. A few examples of SMEFT operators that generate exotic high-multiplicity muon decays are listed in \cref{tab:SMEFT_bounds}, where $\ell$ denotes left-handed lepton doublets, $e$ right-handed leptons, $H$ the Higgs doublet, and $B_{\mu\nu}$ the hypercharge field strength. The operators are of two types: those that lead to muon decays with only electrons and positrons in the final state (ignoring radiative corrections), and those that, in addition, lead to decays involving photons. 

Feynman diagrams for each of the examples in \cref{tab:SMEFT_bounds} have final-state particles flowing directly from the SMEFT operator vertex. In principle, the operators that mediate the decay $\mu\to (me)(n\gamma)$ can also induce the decay $\mu\to ((m+2)e)((n-1)\gamma)$, if one of the photons is off-shell and Dalitz converts into an $e^+e^-$ pair. For example, the  transition magnetic moment operator $(\bar e \sigma^{\mu\nu} \mu) F_{\mu\nu}$ induces the decay $\mu\to e \gamma$, but also the $\mu\to 3e$ decay via $\mu\to e (\gamma^*\to 2e)$. However, the decay rates for such Dalitz conversion processes are suppressed by ${\mathcal O}(\alpha/\pi)\sim {\mathcal O}(10^{\eminus3})$ relative to the original decay. Similarly, the SMEFT operators that induce $\mu \to (me)(n\gamma)$ decays can also mediate $\mu \to (me)((n+1)\gamma)$ decays, where a hard photon is emitted from the electron leg. These processes are likewise suppressed by ${\mathcal O}(\alpha/\pi)$.

Experimentally, the leading sensitivity to a particular SMEFT operator arises from the processes in which all final-state particles attach directly to the SMEFT operator vertex. For example, the $\mu \to 5e$ transition could be induced by a $\mu \to 3e$ point-like SMEFT interaction, followed by QED emission of a $e^+e^-$ pair. The projected sensitivity to ${\mathrm{BR}}(\mu \to 3e)$ is $10^{\eminus16}$~\cite{Ardu:2024bua, Calibbi:2021pyh}, which means that unless the NP is discovered, the $\mu \to 5e$ rate from transitions involving QED radiation will be suppressed well below our benchmark experimental sensitivity for high-multiplicity channels of $\cO(10^{\eminus15})$. Similar conclusions apply to all other high-multiplicity final states.\footnote{The one exception is the low multiplicity decay $\mu \to e \gamma$. Here, the future sensitivities to transition magnetic moment from $\mu\to e$ conversion and $\mu\to 3e$ searches will be comparable to those from $\mu \to e \gamma$ decay \cite{Calibbi:2017uvl}.}

This motivates us to concentrate on local amplitudes induced by SMEFT operators. For each signature listed in \cref{tab:SMEFT_bounds}, we identify a representative SMEFT operator with minimal parametric suppression. First, we select the lowest dimension SMEFT operator capable of generating the process via a single-vertex Feynman diagram. Second, we consider the parametric sizes of contributions: insertions of the Higgs doublet introduce factors of $v_\text{ew}$, and are thus preferred over derivatives, which contribute at most factors of $m_\mu$.\footnote{For SMEFT operators with a chirality flip associated with a Higgs insertion, we also include a suppression by the corresponding SM Yukawa coupling, i.e., $\cC_i\sim y_\mu$, as expected in UV models that explain the SM flavor hierarchies. In this case, the Higgs insertion leads to a suppression $y_\mu v_\text{ew}\sim m_\mu$ and is thus parametrically of the same size as an additional derivative in the operator.}

For each signature, we estimate the sensitivity to the NP scale $\Lambda$ of the most promising operator by imposing $\mathrm{BR}(\mu\to e+X) < 10^{\eminus15}$, representative of future experimental reach on higher-multiplicity modes.\footnote{For some channels, such as ${\color{red}{\mu\to e2\gamma}}$, this may be too optimistic. 
However, if the sensitivity is worse than $\mathrm{BR}(\mu\to e+X)<10^{\eminus15}$, then the NP scales probed will be even lower than those quoted in \cref{tab:SMEFT_bounds}, and thus our conclusions will not change. For instance, the current bound $\mathrm{BR}(\mu\to e2\gamma)<7.2\times10^{\eminus11}$\,\cite{Bolton:1988af} corresponds to a $\Lambda\gtrsim 20\,\gev$.} The branching ratios were computed using the simulation pipeline \texttt{FeynRules}~\cite{Alloul:2013bka} $\to$ \texttt{UFO}~\cite{ Degrande:2011ua} $\to$ \texttt{MadGraph}~\cite{Alwall:2011uj}. The resulting sensitivities are summarized in \cref{tab:SMEFT_bounds}. 

The golden decay channels $\mu\to e \gamma$ and $\mu\to 3e$ proceed through dimension-6 SMEFT operators, and probe physics well beyond the electroweak scale. In contrast, the SMEFT induced high-multiplicity muon decays probe $\Lambda$ that is at most of the weak scale, making the use of SMEFT description itself suspect.\footnote{Note that we have included the muon Yukawa suppression for operators with a chirality flip, and a loop factor $(16\pi^2)^{\eminus1}$ for dipole and Rayleigh operators, as expected from renormalizable UV completions.} Note that some of the decay modes in \cref{tab:exotic_signatures} were omitted from \cref{tab:SMEFT_bounds}, such as ${\color{blue}{\mu\to 7e}}$ and ${\color{blue}{\mu\to 5e\gamma}}$. These require even higher-dimensional SMEFT operators; for example, ${\color{blue}{\mu\to 7e}}$ requires an operator of at least dimension 12, and thus the probed NP scale $\Lambda$ is even lower.

If we include radiative corrections, the high-multiplicity SMEFT operators are also probed by the lower-multiplicity transitions. For example, the SMEFT operator that induces the $\mu \to ne$ decay will also lead to $\mu \to (n-2)e$ decay at 1-loop, via the same type of Feynman diagrams, but closing $e^+$ and $e^-$ legs into a loop. If the experimental sensitivity to both decays is comparable, the two transitions probe similar NP scales, at least parametrically: the $\mu \to ne$ decay is suppressed by a $1/(16\pi^2)^2$ phase space factor compared to the $\mu \to (n-2)e$ decay rate, while the amplitude for  $\mu \to (n-2)e$ decay is suppressed by the $1/(16\pi^2)$ loop factor (and thus the rate by a factor $1/(16\pi^2)^2$). Apart from implicit UV model dependence, possible cancellations and additional factors, such as chirality flips,  which were neglected in the simplified scaling argument above, one might be tempted to conclude that measuring high-multiplicity muon decays is unnecessary. After all, within the SMEFT framework, these processes are effectively already constrained by bounds on $\mu \to e\gamma$, $\mu \to 3e$, and $\mu \to e$ conversion. However, as we will show in the following subsection, this conclusion no longer holds in the presence of light new physics states.

Before we proceed, let us comment on the intermediate mass regime, where the new states $X$ are lighter than the electroweak scale, yet heavy enough that they cannot be produced in muon decays, $m_\mu < m_X < v_{\text{ew}}$. In this regime, SMEFT is no longer applicable, and needs to be replaced by the Low Energy Effective Field Theory (LEFT), in which electroweak symmetry is broken and the remaining gauge group is $SU(3)_c\times U(1)_{\text{em} }$~\cite{Jenkins:2017jig}. For muon decays, the relevant part of LEFT consists of the higher-dimensional operators built from muon, electron, neutrino, and photon fields. The effects of $X$ are captured by local operators generated when $X$ is integrated out. We will encounter an explicit UV realization of this type in \cref{sec:model_off_shell_ALP}, leading to  ${\color{red}{\mu\to e2\gamma}}$ decays.

\subsection{SMEFT${}_X$}
\label{sec:EFT:lightNP}

The analysis above is drastically modified if there are light new physics states to which the muon can decay. We consider the simplest possibility, where we extend the SMEFT by a single additional light NP state $X$, leading to the light-particle EFT framework that we refer to as SMEFT${}_X$. The key difference with respect to the SMEFT is that now the high-multiplicity final state can appear due to the decays of $X$. Below, we examine four concrete SMEFT${}_X$ realizations, each involving a single new SM-singlet field: {\em (i)} a scalar $\mathcal{S}$ or {\em (ii)} a pseudoscalar $\mathcal{P}$, {\em (iii)} a fermion $N$, and {\em (iv)} a vector $V$. The corresponding effective operators and the scales they can probe are summarized in \cref{tab:Leff:light}, with the sensitivities extracted under the assumption $\mathrm{BR}(\mu \to e + X) \sim 10^{\eminus15}$. We provide only order-of-magnitude estimates, cross-checked with \texttt{MadGraph}, which remain valid except near the kinematic endpoints, where the precise value of $m_X$ becomes important. 

From the results in \cref{tab:Leff:light} we see that the same type of $\mu \to (ne)(m\gamma)$ decays now probe much higher UV scales compared to SMEFT (see \cref{tab:SMEFT_bounds}). The reason is that the $\mu\to n X$ decays are now mediated by the operators of lower dimension than those appearing in the SMEFT description. Another key difference relative to SMEFT is that the relations to lower multiplicity transitions, when including radiative corrections, get modified: diagrams obtained by closing the $e^+ e^-$ legs into loops are additionally suppressed by small couplings of $X$ to electrons. Closing $X$ legs into a loop may instead still lead to parametrically relevant, though UV model-dependent, constraints. 

Results in \cref{tab:Leff:light} assume that $X$ particles decay into either photons or electrons within the detector volume. This implies sufficiently large couplings, so that the models face also constraints due to collider, low-energy precision, beam-dump, and astrophysical searches. For instance, prompt $X \to e^+e^-$ decays, where $X = \mathcal{S}, \mathcal{P}, V$, are allowed, but typically require $m_X \gtrsim 10\,\mathrm{MeV}$, see, e.g., Refs.~\cite{Heeck:2017xmg, Greljo:2021npi, DiLuzio:2025ojt}. Below, we give further details on each of the four SMEFT${}_X$ realizations.

\begin{table}[t]
\centering

\centering
\renewcommand{\arraystretch}{1.2}
\setlength{\tabcolsep}{18pt}  
\scalebox{1}{
\begin{tabular}{c@{\hspace{1.85cm}}c}
\midrule\midrule
${\bm{\cL_{\sscript{eff}}\supset}}$
&${\bm{\Lambda\,[\gev]}}$
\\[1pt]
\midrule
$\frac{y_\mu}{\Lambda}(\bar\ell_2 H e_1)\,\cS$
&$\cO(10^{14})$
\\[3pt]
$\frac{y_\mu}{\Lambda^2}(\bar\ell_2 H e_1)\,\cS^2$
&$\cO(10^{6})$ 
\\[3pt]
$\frac{y_\mu}{\Lambda^3}(\bar\ell_2 H e_1)\,\cS^3$
&$\cO(10^3)$
\\[3pt]
$\frac{y_\mu}{\Lambda^4}(\bar\ell_2 H e_1)\,\cS^4$
&$\cO(10^2)$
\\[3pt]
$\frac{1}{\Lambda^{3}}(\bar f_2\gamma^\mu f_1)(\bar f_1\gamma_\mu f_1)\,\cS$
&$\cO(10^3)$
\\[3pt]
$\frac{1}{\Lambda^{4}}(\bar f_2\gamma^\mu f_1)(\bar f_1\gamma_\mu f_1)\,\cS^2$
&$\cO(10^2)$
\\[3pt]
$\frac{1}{16\pi^2}\frac{y_\mu}{\Lambda^3}(\bar\ell_2\sigma^{\mu\nu}e_1)HB_{\mu\nu}\,\cS$
&$\cO(10^3)$
\\[3pt]
$\frac{1}{16\pi^2}\frac{y_\mu}{\Lambda^4}(\bar\ell_2\sigma^{\mu\nu}e_1)HB_{\mu\nu}\,\cS^2$
&$\cO(50)$
\\[3pt]
\midrule[0.1pt]
$\frac{y_\mu}{\Lambda}(\bar\ell_2 H e_2)\,\cS$
&$\cO(10^7)$
\\[3pt]
$\frac{y_\mu}{\Lambda^2}(\bar\ell_2 H e_2)\,\cS^2$
&$\cO(10^2)$
\\[3pt]
\midrule
$\frac{1}{\Lambda} (\bar f_2\gamma^\mu f_1) \partial_\mu \mathcal P$
&$\cO(10^{14})$
\\[3pt]
$\frac{1}{16\pi^2}\frac{1}{\Lambda^3}(\bar f_2\gamma_\nu f_1)B^{\mu\nu} \partial_\mu \mathcal P$
&$\cO(10^3)$
\\[3pt]
$y_\mu\frac{1}{\Lambda^5}(\bar f_2\gamma^\mu f_1)(\bar\ell\, H e) \partial_\mu \mathcal P$
&$\cO(50)$
\\[3pt]
$\frac{1}{16\pi^2}\frac{1}{\Lambda^5}(\bar f_2\gamma^\mu f_1)B_{\rho\sigma}B^{\rho\sigma} \partial_\mu \mathcal P$
&$\cO(10)$
\\[3pt]
$\frac{y_\mu}{\Lambda^4}  (\bar\ell_2 H e_1) (\partial_\mu \mathcal P)^2$
&$\cO(10^2)$
\\[3pt]
\midrule
$\frac{1}{\Lambda^2}(\bar\ell_2\gamma^\mu \ell_1)(\bar N\gamma_\mu N)$
&$\cO(10^6)$
\\[3pt]
$\frac{y_\mu}{\Lambda^6}(\bar\ell_2 H e_1)(\bar N\gamma_\mu N)(\bar N\gamma^\mu N)$
&$\cO(1)$
\\[3pt]
\midrule
$\frac{1}{\Lambda^2}(\Phi^\dag i\overleftrightarrow D_\mu \Phi)(\bar f_2\gamma^\mu f_1)$
&$\cO(10^7)$
\\[3pt]
$\frac{y_\mu}{\Lambda^4}(D_\mu\Phi)^\dag(D^\mu\Phi)(\bar\ell_2 He_1)$
&$\cO(10^3)$
\\[3pt]
\midrule\midrule
\end{tabular}
}
\caption{Representative SMEFT${}_X$ operators contributing to the exotic muon decays with light SM singlet mediators produced on-shell: scalars ($\cS$), pseudoscalars ($\mathcal P$), fermions ($N$), and vectors ($V$). For each flavor-violating operator (and likewise for $\cS$, shown for two representative flavor-diagonal couplings to muon), we estimate the lower bound on the effective scale $\Lambda$ corresponding to $\mathrm{BR}(\mu \to e+X) < 10^{\eminus15}$. The operator normalization follows the conventions adopted in \cref{tab:SMEFT_bounds} and, where relevant, $f=\{\ell,e\}$. The light vector $V$ is assumed to be a gauge boson of a spontaneously broken dark $U(1)_d$ with $\Phi$ the dark Higgs.  See \cref{sec:EFT:lightNP} for more details.}
\label{tab:Leff:light}
\end{table}

\vspace{0.2cm}
\noindent
{\textbf{Light scalar EFT.}} In this example, SMEFT is supplemented by a light scalar $\mathcal{S}$. To simplify the discussion, let us first assume that the $\mu \to e + n\,\mathcal{S}$ transitions arise from Yukawa-like SMEFT operators built out of the SM bilinears $\bar\ell_2 H e_1$ or $\bar\ell_1 H e_2$, and a fixed number of scalar field insertions. Specifically, out of the full SMEFT${}_X$ Lagrangian we focus on the operators
\beq\label{eq:light_scalar_EFT_Yukawa_type}
 \cL_{\sscript{eff}}\supset \frac{\cC^{(n)}_\cS}{\Lambda^{n}}\big(\bar \ell_2 H e_1)\cS^n+\hermc\,.
\eeq
The operators that differ by the chiralities of electron and muon fields  ($1\leftrightarrow 2$ indices exchanged) lead to the same phenomenology, and can be ignored for simplicity.

The naive dimensional analysis (NDA) estimate for the $\mu \to e + n\,\mathcal{S}$  branching ratio is 
\begin{equation}
    \mathrm{BR}(\mu\to e+n\cS)\sim 
    \frac{y_\mu^2\, v_\text{ew}^6}{(16\pi^2)^{n-2}}\frac{m_\mu^{2n-6}}{\Lambda^{2n}}\,,
\end{equation}
where we assumed that $\cC_\cS^{(n)}\sim y_\mu$, which would be the generic expectation in many UV completions. Assuming that $\mathcal{S} \to 2e$ or $\mathcal{S} \to 2\gamma$ are prompt, with nearly unit branching ratio, the above estimate for $\mathrm{BR}(\mu\to e+n\,\cS)$ also gives the branching ratio for the corresponding high-multiplicity muon decay, $\mu \to (ne) (m\gamma)$.

Taking as a benchmark the experimental sensitivity to such exotic decays to be $\mathrm{BR}(\mu \to e + n\,\cS) \sim 10^{\eminus15}$, we derive the corresponding bounds on the suppression scale $\Lambda$, as given in the upper part of \cref{tab:Leff:light}.
We observe that for $n=1,2,3$, corresponding to $\mu\to 3e, 5e, 7e$ decays for $\cS\to 2e$, the inferred effective scales exceed the electroweak scale. This motivates restricting the analysis to a finite number (at most 4) of light $\mathcal{S}$ particles produced in the $\mu \to e+n\,\mathcal S$ transition. 
In practice, this limitation may also be driven by experimental considerations. For example, for $n = 4$ with each $\mathcal{S}$ decaying via $\mathcal{S} \to 2e$, the exotic muon decay is $\mu \to 9e$. In this case, each electron and positron has on average a momentum of only $\mathcal{O}(10\,\mev)$, which is right at the detection threshold of the Mu3e experiment~\cite{Hesketh:2022wgw}.\footnote{The detection thresholds depend on the experimental setup. For instance, in the CrystalBox experiment, all the energy released in the form of electrons/positrons and photons got deposited in the calorimeter \cite{Bolton:1988af}.}

Another class of operators we consider consists of dimension-6 charged Lepton Flavor Violating (cLFV) four-fermion interactions dressed with multiple insertions of light scalar field $\cS$:
\begin{equation}\label{eq:light_scalar_EFT_SMEFT_cLFV_times_scalar}
    \begin{split}
        \cL_{\sscript{eff}}&\supset \frac{\cC_\cS^{(n),1}}{\Lambda^{2+n}}(\bar\ell_2\gamma^\mu\ell_1)(\bar\ell_1\gamma_\mu\ell_1)\cS^n
        \\
       &+\frac{\cC_\cS^{(n),2}}{\Lambda^{2+n}}(\bar\ell_2\gamma^\mu\ell_1)(\bar e_1\gamma_\mu e_1)\cS^n
        \\&
        +\frac{\cC_\cS^{(n),3}}{\Lambda^{2+n}}(\bar e_2\gamma^\mu e_1)(\bar e_1\gamma_\mu e_1)\cS^n+\hermc
        \,.
    \end{split}
\end{equation}
These operators induce $\mu \to 3e + n\,\cS$ decays, with the NDA estimates for the branching ratios given by 
\begin{equation}
    \mathrm{BR}(\mu\to 3e+n\,\cS)\sim \frac{v_\text{ew}^4}{(16\pi^2)^n}\frac{m_\mu^{2n}}{\Lambda^{4+2n}}\,,
\end{equation}
where we assumed that $\cC_\cS^{(n),i}\sim1$. The corresponding values of the effective NP scale $\Lambda$ that can be probed for different values of $n$ are summarized in \cref{tab:Leff:light}. Notably, we observe that already for $n \geq 3$, the inferred sensitivity to $\Lambda$ drops below the electroweak scale. This implies that for $n=1,2$ it is straightforward to write down UV completions in which all NP states are heavy (apart from $\cS$), and thus not excluded by direct searches. In contrast, for $n\geq 3$, the UV completions very likely require additional light NP states.

Another distinct class of SMEFT${}_X$ operators is given by dimension-6 cLFV dipole-like terms with $n$ scalar insertions,
\begin{equation}
\label{eq:diopole:Sn}
     \cL_{\sscript{eff}}\supset \frac{1}{16\pi^2}\frac{\cC_\cS^{(n)}}{\Lambda^{2+n}}(\bar\ell_2\sig^{\mu\nu}e_1)HB_{\mu\nu}\,\cS^n+\hermc\,.
\end{equation}
These operators lead to $\mu \to e\gamma + n\,\mathcal{S}$ decays  with an estimated branching ratio of 
\begin{equation}
    \mathrm{BR}(\mu\to e\gamma\, n\,\cS)\sim \frac{y_\mu^2\, v_\text{ew}^6}{(16\pi^2)^{n+1}}\frac{m_\mu^{2n-2}}{\Lambda^{4+2n}}\,,
\end{equation}
where we assumed that $\cC_\cS^{(n)}\sim y_\mu$. 

\cref{tab:Leff:light} summarizes the values of the scale $\Lambda$ that can be probed for $n=1,2$. As with scalar-dressed four-fermion operators in \cref{eq:light_scalar_EFT_SMEFT_cLFV_times_scalar}, the effective scales probed for dipole-type operators are generally lower than those for the Yukawa-type operators given by \cref{eq:light_scalar_EFT_Yukawa_type}. This is true both when the value of $n$ is kept fixed in the comparisons and when comparing decays with the same final state multiplicities. The differences in the effective scale reach between the three types of operators are due to both the suppression of the corresponding SMEFT${}_X$ operators and the strong phase-space penalty associated with the high-multiplicity final states that arise directly from the SMEFT${}_X$ operator vertex. For instance, the $\mu\to 5e$ decay induced by the $\bar \ell_2 He_1\cS^2$ operator proceeds via three-body decay $\mu\to e2\cS$, where $\cS\to2e$, and is due to a dimension-6 SMEFT${}_X$ operator. In contrast, if $\mu\to 5e$ decay is induced by the $(\bar f_2\gamma^\mu f_1)(\bar f_1\gamma_\mu f_1)\,\cS$ operator, the $\mu\to 5e$ signature arises from a four-body decay $\mu\to 3e\cS$, while the SMEFT${}_X$ operator is of dimension 7. 

The dipole-type operators in \cref{eq:diopole:Sn} face another important set of constraints: at 1-loop the electroweak RG running mixes $\bar \ell_2 \sigma^{\mu \nu} e_1 H B_{\mu \nu} \cS^n$ into Yukawa-like operators,  $\bar \ell_2 e_1 H^3 \cS^n$. Setting the Higgs field to its vacuum expectation value (VEV), these operators induce $\mu \to e +n\cS$ decays, similarly to the operators in \cref{eq:light_scalar_EFT_Yukawa_type}, but now with coefficients that are additionally suppressed by a factor $(v_\text{ew}/\Lambda)^2 (1/16\pi^2)$. Despite this suppression, the bounds on $\mu \to e +n\,\cS$ branching ratios lead to more stringent constraints on $\Lambda$ than do the bounds from the channels with an extra photon.

As a final example of SMEFT${}_X$ operators with $\cS$ as the light NP state, we consider scenarios in which $\cS$ is emitted from a SM lepton leg. At the same time, the muon decays proceed via the SM Fermi interaction~\cite{Knapen:2024fvh, Knapen:2023iwg}, leading to the $\mu\to n\,\cS+e\nu\bar\nu$ signature. The emission of $\cS$ is mediated by \textit{flavor-conserving} operators, such as 
\begin{equation}
     \cL_{\sscript{eff}}\supset
\frac{y_\mu}{\Lambda}(\bar\ell_2 He_2)\cS^n +\hermc\,,
\end{equation}
where we included the expected suppression by the muon Yukawa. The expected sensitivities to the effective NP scale $\Lambda$ are collected in \cref{tab:Leff:light}. For instance, for the dimension-5 operator $(\bar\ell_2 He_2)\cS$ one could reach $\Lambda \sim \cO(10^7\,\mathrm{GeV})$,  demonstrating that even flavor-conserving operators emitting light states from external legs can probe very high scales.

\vspace{0.2cm}
\noindent
{\textbf{Light pseudoscalar EFT.}} Next, let us consider the possibility of a light pseudoscalar, $\cP$. We assume that $\cP$ is a pseudo-Nambu-Goldstone Boson (pNGB), and thus only has derivative couplings to the SM currents (apart from the anomaly terms). Importantly, the interaction Lagrangian for  $\cP$  is invariant under the shift symmetry  $\cP\to \cP+\phi$, with $\phi$ an arbitrary constant parameter. This shift symmetry is broken only by the nonzero $\cP$ mass. Since the limit of a massless $\cP$ corresponds to an enhanced symmetry point, the pNGB  $\cP$ can be naturally light. By contrast, the scenarios featuring a light scalar $\cS$, which we discussed above, suffer from a hierarchy problem.

The SMEFT${}_X$ interactions coupling pNGB to the SM start at dimension-5~\cite{MartinCamalich:2020dfe,Calibbi:2020jvd,Bauer:2020jbp, Bauer:2021mvw, Galda:2021hbr, Greljo:2024evt, Alda:2025uwo}. Among the SMEFT${}_X$ operators, we are specifically interested in those that induce exotic muon decays, with the two lowest-dimension operators involving a single insertion of $\cP$ given by
\begin{equation}
\label{eq:cP:Leff}
    \cL_{\sscript{eff}} \supset \frac{\partial_\mu \mathcal P}{\Lambda}(\bar f_2\gamma^\mu f_1) + 
\frac{1}{16\pi^2}\frac{\partial_\mu \mathcal P}{\Lambda^3}(\bar f_2\gamma_\nu f_1)B^{\mu\nu} + \dots\,.
\end{equation}
In \cref{tab:Leff:light} we also list two examples of dimension-9 operators as well as a Yukawa-like interaction involving two $\cP$ fields. 

For each of the benchmark operators, \cref{tab:Leff:light} collects the expected sensitivity to the effective NP scale $\Lambda$. The results are closely related to the ones we obtained for the processes involving light scalar $\cS$; in the NDA estimates, the derivative on $\cP$ contributes a factor of $m_\mu$ since this is the typical scale in the muon decays, which is the same suppression factor $y_\mu v$ encountered in the corresponding operators involving $\cS$. In other words, the muon decays induced by the $\partial_\mu \mathcal P (\bar f_2\gamma^\mu f_1)$ and $y_\mu (\bar\ell_2 H e_2)\,\cS$ operators are expected to probe similar effective scale $\Lambda$, and similarly for $(\bar f_2\gamma_\mu f_1)B^{\mu\nu} \partial_\nu \mathcal P$ and $y_\mu (\bar\ell_2\sigma^{\mu\nu}e_1)HB_{\mu\nu}\,\cS$, as evident from \cref{tab:Leff:light}. 

However, the similarity between the phenomenology of SMEFT${}_X$ for $\cP$ and $\cS$ ends when higher insertions of $\cP$ field are considered. While each insertion of $\cP$ comes in the form $\partial_\mu \cP$ and thus with effective power suppression by $m_\mu$, this is not the case for $\cS$. This means that while $\mu\to e\cS$ and $\mu\to e\cP$ decays probe similar NP scale $\Lambda$, the decay $\mu\to e+2 \cP$ probes  $\Lambda$ of only the order weak scale, in stark contrast to $\mu\to e+2 \cS$, which probes an effective scale many orders of magnitude above the weak scale. 

In short, SMEFT${}_X$ with light pNGB $\cP$ can lead to high multiplicity muon decays only if the effective NP scale $\Lambda$ is relatively low. However, this conclusion is modified if pNGBs arise from a non-Abelian symmetry-breaking pattern: in this case, multiple insertions of $\cP$ fields can still be suppressed by only a single derivative, while preserving the shift symmetry~\cite{Bigaran:2025uzn}. More specifically, operators of the form $\cP_1 \ldots \cP_{n-1}\partial_\mu \mathcal P_n (\bar f_2\gamma^\mu f_1)$  are then allowed by the shift symmetry. In this setup, the NDA estimates of sensitivity to NP scales closely resemble the discussion of the case of SMEFT${}_X$ for $\cS$.

\vspace{0.2cm}
\noindent
{\textbf{Light fermion EFT.}} A qualitatively different set of signatures in exotic muon decays is obtained if we extend SMEFT by a light SM-singlet fermion $N$. In this case, the leading cLFV interactions involve two insertions of $N$ field, and are due to dimension-6 operators of the form
\beq
 \cL_{\sscript{eff}}\supset\frac{\cC_N^{(1)}}{\Lambda^2}(\bar\ell_2\gamma^\mu \ell_1)(\bar N\gamma_\mu N)+\hermc\,.
\eeq 
In this case $\mathrm{BR}(\mu \to e N N) \sim 10^{\eminus15}$ probes the effective scales $\Lambda\sim \cO(10^6\,\gev)$ (see \cref{tab:Leff:light}). For higher-dimension operators, this scale quickly becomes very low. For instance, dimension-7 operator $y_\mu (\bar\ell_2 H \ell_1)(\bar N  N)$ only probes $\Lambda \sim {\mathcal O}(10^3)$. In contrast, operators containing an additional $\bar N N$ pair are even further suppressed, failing to probe meaningful NP scales.

An important consideration is the decay pattern of $N$: for it to be experimentally detected, it must decay into electrons and/or photons within the detector volume. This cannot be achieved via the renormalizable Heavy-Neutral-Lepton (HNL) portal ${\cal L}_\sscript{HNL}=y_N (\bar\ell^c HN)$, given the stringent experimental bounds~\cite{Fernandez-Martinez:2023phj}. Similarly, dimension-6 transition magnetic moment interaction $(\bar \ell^c \sigma^{\mu\nu} N)H B_{\mu\nu}$ induces $N \to \nu \gamma$ decays; however, stringent constraints on this type of interaction imply that such decays can only occur with highly displaced vertices~\cite{Brdar:2020quo}. A more promising possibility, that can lead to prompt decays, is $N$ decaying into another light BSM neutral fermion (potentially dark matter), along with electrons and/or photons. This possibility is explored in \cref{sec:model_dark_higgs} (see also alternative possibilities, such as examples in Refs.~\cite{Ballett:2019pyw, Bansal:2022zpi}).

\vspace{0.2cm}
\noindent
{\textbf{Light vector EFT.}} Next, let us consider the possibility that the light NP state in SMEFT${}_X$ is a vector $V$. We restrict our discussion to the simplest possibility, namely that $V$ is a gauge boson of a dark $U(1)_d$, spontaneously broken by the VEV of a dark Higgs, denoted by $\Phi$. The SM fields are assumed to be neutral under $U(1)_d$.

In addition, the renormalizable kinetic mixing,
\begin{equation}
\mathcal L \supset -\frac{\varepsilon}{2 c_W} B_{\mu \nu} V^{\mu \nu}\,,   
\label{eq:kin:mixing}
\end{equation}
with $c_W$ the cosine of the Weinberg angle,
induces couplings of $V$ to the SM, proportional to the electromagnetic current, ${\cal L}_\sscript{int}\supset \varepsilon V_\mu J_\sscript{em}^\mu$, where $J_\sscript{em}^\mu=\sum_f e Q_f\bar f \gamma^\mu f$, with $Q_f$ the electric charge of a SM fermion $f$. These interactions are flavor universal, and induce $V\to e^+e^-$ decays as the dominant decay channel for vector boson $V$, in the region of parameter space of interest.

Flavor-violating interactions of the vector boson $V$ arise through higher-dimensional operators. The leading contribution is due to the dimension-6 operator
\begin{equation}\label{eq:light_vec_dim6}
	\cL_{\sscript{eff}}\supset \frac{\cC_V^{(1)}}{\Lambda^2}(\Phi^\dag i \overset{\text{\footnotesize$\leftrightarrow$}}{D}_\mu \Phi)(\bar f_2\gamma^\mu f_1)+\hermc\,,
\end{equation}
where $D_\mu$ is the covariant derivative of $U(1)_d$, so that $D_\mu \Phi=(\partial_\mu -i g_V V_\mu)\Phi$, and $(\Phi^\dag i \overset{\text{\footnotesize$\leftrightarrow$}}{D}_\mu \Phi)=\Phi^\dag i \partial_\mu \Phi -  i \partial_\mu\Phi^\dag \Phi + 2g_V V_\mu \Phi^\dag   \Phi$. Here,  $g_V$ is the $U(1)_d$ gauge coupling, while the mass of the gauge boson $V$  is given by $m_V=g_V v_\Phi$.

Once the dark Higgs $\Phi$ acquires a vacuum expectation value $v_\Phi$, the dimension-6 operator in \cref{eq:light_vec_dim6} induces an interaction between $V_\mu$ and the flavor violating lepton current:
\beq
    \cL_V\supset 2 \cC_V^{(1)} g_V\frac{v_\Phi^2}{\Lambda^2}V_\mu (\bar f_2 \gamma^\mu f_1)+\hermc\,,
\eeq
which is then responsible for  $\mu\to e V$ decays. As the numerical example, upon setting $g_V\sim \cC_V^{(1)}\sim {\mathcal O}(1)$ and $v_\Phi \sim m_\mu$, the NDA estimate for the $\mu \to e V$ branching ratio reads
\begin{equation}
    \mathrm{BR}(\mu\to e V)\sim \frac{1}{8\pi\Gamma_\mu}\frac{m_\mu^5}{\Lambda^4}\,,
\end{equation}
where $\Gamma_\mu$ is the SM decay width for the muon. Adopting $\mathrm{BR}(\mu\to e(X\to 2e))<10^{\eminus15}$ as the experimental sensitivity, gives the effective NP scale reach of $\Lambda\sim\cO(10^7\,\gev)$, quoted in \cref{tab:Leff:light}.

The $\mu\to e VV$ decays can accordingly be induced by the dimension-8 effective operator
\begin{equation}\label{eq:light_vec_dim8}
	\cL_{\sscript{eff}}\supset \frac{\cC_V^{(2)}}{\Lambda^4}(D_\mu\Phi)^\dag(D^\mu\Phi)(\bar\ell_2 He_1)+\hermc\,.
\end{equation}
As for the dimension-6 case, taking $g_V\sim {\mathcal O}(1)$, $v_\Phi \sim m_\mu$ and $ \cC_V^{(2)}\sim y_\mu$, as usual for the chirality flipping operators, the NDA estimate for the $\mu\to e VV\to 5e$ branching ratio becomes
\beq
    \mathrm{BR}(\mu\to eVV)\sim\frac{1}{512\pi^3}\frac{1}{\Gamma_\mu}\frac{m_\mu^9}{\Lambda^8}\,.
\eeq
A benchmark bound $\mathrm{BR}(\mu\to e\,2 (X\to 2e))<10^{\eminus15}$ would then translate to a sensitivity to the effective NP scale of $\Lambda\sim\cO(10^3\,\gev)$.

The previous discussion can be modified in various ways. For instance, as mentioned above, we assumed that the flavor diagonal couplings of $V_\mu$ are due to kinetic mixing (see \cref{eq:kin:mixing}), while the SM fermions do not carry $U(1)_d$ charges. One could instead extend the SM by an anomaly-free $U(1)$ with flavor non-universal gauge charges for the SM fermions; examples include the $U(1)_{\mu-e}$ gauge group as well as other arrangements~\cite{Greljo:2021npi, Greljo:2022dwn, Allanach:2020zna, Costa:2020dph, Smolkovic:2019jow}. In this case, the leading exotic muon decay is either $\mu\to 5e$ or $\mu\to 3e$, depending on whether the dark Higgs is light enough to be produced in muon decays, see~Appendix~\ref{sec:AppU(1)mu-e} for a more comprehensive discussion. Another possibility is that the vector $V_\mu$ gets emitted (due to its flavor diagonal couplings) from either the muon or the electron leg in the SM Michel decay $\mu\to e \nu\bar \nu$. This occurs both in the $U(1)_{\mu-e}$ model and for the kinetic-mixing portal (dark photon) model. Such emissions are phenomenologically relevant if the flavor diagonal couplings of $V_\mu$ are not too small~\cite{Knapen:2023iwg}. The phenomenology of dark vectors can also be much more complex; they could, for instance, be part of a non-Abelian gauge symmetry, so that gauge-invariant SMEFT${}_X$ operators then require at least two insertions of $V_{\mu\nu}$. The vector $V_\mu$ could also be part of a more complex dark sector, and not necessarily even the lightest state, as in composite dark sector models (see, e.g., Ref.~\cite{Berlin:2018tvf}).

\section{Benchmark New Physics Models}
\label{sec:models}

The EFT setup discussed so far does not capture phenomenology beyond muon decays. To put the sensitivity of rare muon decays into a broader perspective of other new physics searches, we present several benchmark models where one or more of the exotic muon decays listed in \cref{tab:exotic_signatures} serve as competitive signatures.

\subsection{Model I: Flavor-Protected Scalar}
\label{sec:modelI}

The first benchmark model is a UV completion of the SMEFT${}_X$ effective interactions\footnote{Note the change in the assumed chiral structure between the operators in \cref{eq:opsModelI} and those shown in \cref{tab:Leff:light}. This does not affect any of the phenomenology we consider.}
\begin{equation}\label{eq:opsModelI}
\mathcal{L} \supset \frac{\cC_{\cS}^{(3)}}{\Lambda^3}\,\bar\ell_{1} H e_{2}\, \cS^3 + \frac{\cC_{e\cS}}{\Lambda}\bar\ell_{1} H e_{1} \cS+ \hermc\,,
\end{equation}
where $\cS$ is an SM singlet. The two operators induce the $\mu\to 7e$ signature via a decay chain $\mu\to e\, 3\cS$, $\cS\to 2e$. Before we present the UV model for the two interactions, let us first refine the NDA estimates for the expected reach on the effective scale $\Lambda$ presented in the third row in \cref{tab:Leff:light}.

\begin{figure}[t]
	\centering
	\includegraphics[width=0.5\textwidth]{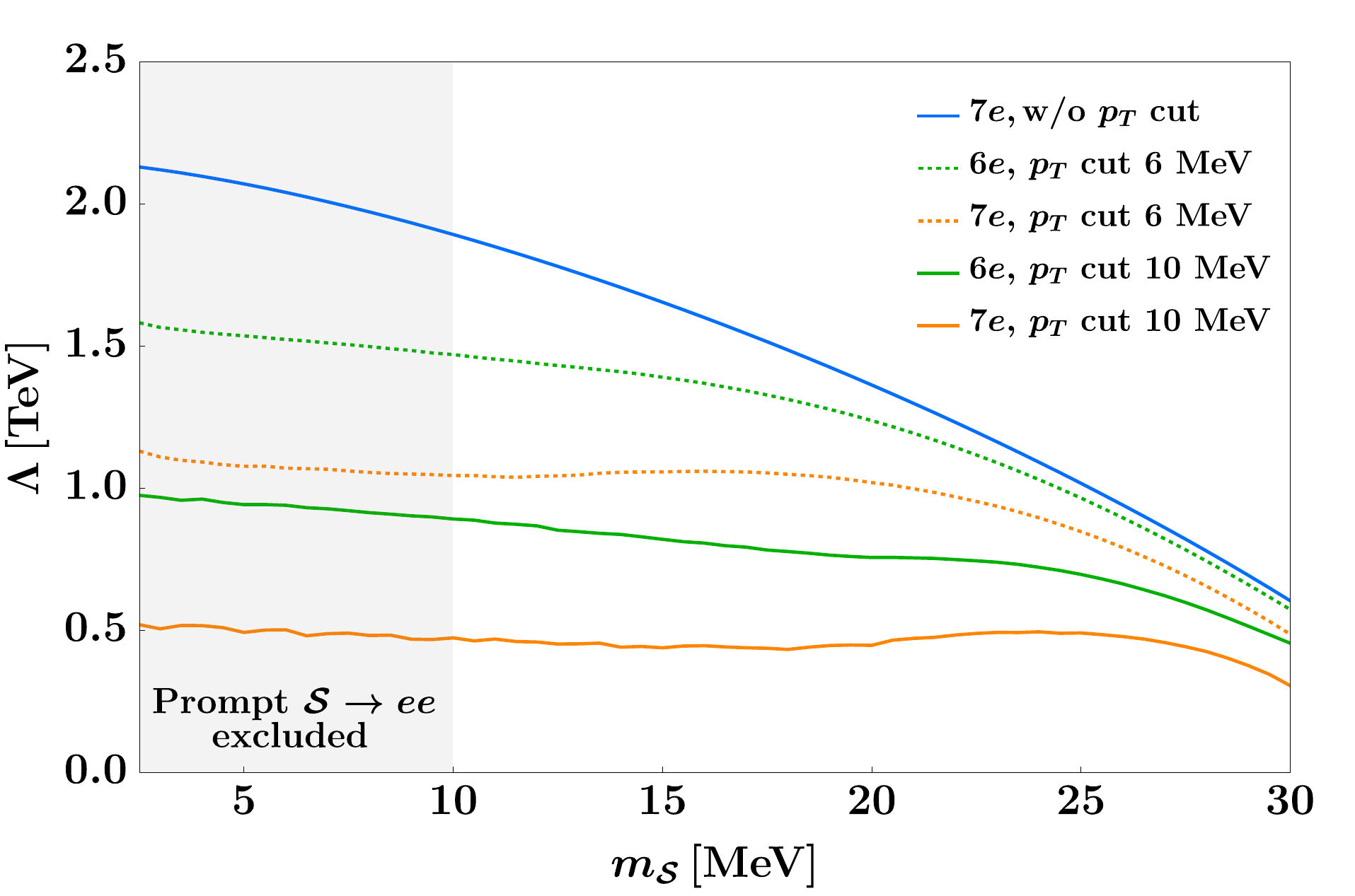}
    \caption{Sensitivity to the effective UV scale $\Lambda$ in \cref{eq:opsModelI}, setting $\cC_\cS^{(3)}=y_\mu$ from decay $\mu \to e + 3\,\mathcal{S}$, followed by $\mathcal{S} \to 2e$, yielding a 7-electron final state. The curves show the values of $\Lambda$ that yield $\mathrm{BR} \times \mathrm{acceptance} = 10^{\eminus15}$ as a function of the scalar mass $m_\mathcal{S}$, for different kinematical cuts, requiring either 6 or 7 electrons and positrons above the kinematic threshold. The shaded region, $m_\cS\lesssim10\,\mev$, is excluded by the searches at beam-dump experiments~\cite{Heeck:2017xmg}. 
    }
    \label{fig:7e_acceptance}
\end{figure}

As in \cref{tab:Leff:light}, we set $\cC_{\cS}^{(3)}=y_\mu$ and assume that $\cC_{e\cS}$ is large enough so that $\cS\to 2e$ decays are prompt. \cref{fig:7e_acceptance} then gives the expected reach on $\Lambda$, as a function of $\cS$ mass,  $m_\cS \in [2.5,30.0]\,\mev$, assuming zero backgrounds and assuming that the experimental reach for exotic muon decay, including kinematic acceptances, is $\mathrm{BR}\times \mathrm{acceptance} = 10^{\eminus15}$.   Using \texttt{MadGraph} simulation of the signal, we require that either six ($6e$, dashed lines) or all seven ($7e$, solid lines) electrons and positrons from the decay chain satisfy a $p_T$ threshold of $p_T > 10 \,(6,0)\,\mev$, shown as orange (green, blue) lines. We see that the $\Lambda$ reach is fairly sensitive to the $p_T$ threshold. Furthermore, the $6e$ signal is in general more competitive, unless the $p_T$ threshold is assumed to be very small. We have also verified that the $5e$ signature yields comparable sensitivity in the limit of vanishingly small background, though in this case the accidental backgrounds must be carefully considered for a realistic projection~\cite{Hostert:2023gpk}. In general, the reach for the effective NP scale is $\Lambda\sim {\mathcal O}(1\,\tev)$, in agreement with the result quoted in \cref{tab:Leff:light}.

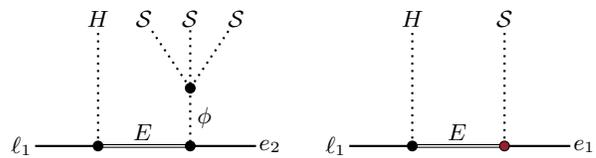
\begin{figure}[t]
\centering
	\begin{tikzpicture}[baseline=(a1.base), scale=0.35, transform shape]
			\begin{feynman}
			\vertex (a1) {\(\scalebox{2.8}{$\ell_1$}\)};
			\vertex [right = 2.9cm of a1](a2);
			\vertex [right = 3.5cm of a2](a3);
			\vertex [right = 2.5cm of a3](a4) {\(\scalebox{2.8}{$e_2$}\)};
			\vertex [above = 4.4cm of a2](a5) {\(\scalebox{2.8}{$H$}\)};
			\vertex [above = 2.2cm of a3](a6);
			\vertex [above = 2.2cm of a6](a7) {\(\scalebox{2.8}{$\cS$}\)};
			\vertex [right = 1.75cm of a7](a7r) {\(\scalebox{2.8}{$\cS$}\)};
			\vertex [left = 1.75cm of a7](a7l) {\(\scalebox{2.8}{$\cS$}\)};
			\vertex [right = 1.7cm of a2](aE);
			\vertex [above = 0.1cm of aE](aEup) {\(\scalebox{2.8}{$E$}\)};
			\vertex [right = 2.5cm of a4](a1n) {\(\scalebox{2.8}{$\ell_1$}\)};
			\vertex [right = 2.9cm of a1n](a2n);
			\vertex [right = 3.5cm of a2n](a3n);
			\vertex [right = 2.5cm of a3n](a4n) {\(\scalebox{2.8}{$e_1$}\)};
			\vertex [above = 4.4cm of a2n](a5n) {\(\scalebox{2.8}{$H$}\)};
			\vertex [above = 2.2cm of a3n](a6n);
			\vertex [above = 2.2cm of a6n](a7n) {\(\scalebox{2.8}{$\cS$}\)};
			\vertex [right = 1.7cm of a2n](aEn);
			\vertex [above = 0.1cm of aEn](aEupn) {\(\scalebox{2.8}{$E$}\)};
			\diagram*{
			(a1) -- [plain, thick] (a2) --[double](a3)--[plain, thick](a4),
			(a2)--[dotted, thick](a5),
			(a7)--[dotted, thick](a6)--[dotted, thick, edge label=\(\scalebox{2.8}{\,$\phi$}\)](a3),
			(a6)--[dotted, thick](a7r),
			(a6)--[dotted, thick](a7l),
			(a1n) -- [plain, thick] (a2n) --[double](a3n)--[plain, thick](a4n),
			(a2n)--[dotted, thick](a5n),
			(a3n)--[dotted, thick](a7n)
			};
			\draw[fill=black] (a2) circle(1.85mm);
			\draw[fill=black] (a3) circle(1.85mm);
			\draw[fill=black] (a6) circle(1.85mm);
			\draw[fill=black] (a2n) circle(1.85mm);
			\draw[fill=red!90!gray] (a3n) circle(1.85mm);
			\end{feynman}
	\end{tikzpicture}
    \caption{Feynman diagrams generating the two operators in \cref{eq:opsModelI} in the  flavor-protected scalar model. The red dot indicates the suppression by the small spurion $\varepsilon_\mu$. See \cref{sec:modelI} for details. \label{fig:modelI}}
\end{figure}

Next, let us move to the UV complete model that generates the two SMEFT${}_X$ interactions in \cref{eq:opsModelI}.  The minimal UV completion requires two additional fields beyond $\cS$: a SM singlet complex scalar $\phi$ and a vector-like charged lepton $E_{L, R} \sim ({\bf 1},{\bf 1})_{\eminus1}$ that has the same quantum numbers as the right-handed SM charged lepton.\footnote{Here and in the rest of this section the notation for the SM gauge representation is $(SU(3)_c, SU(2)_L)_{U(1)_Y}$.} By definition, the $E_R$ are those fields that have the vector-like mass term, and the SM right handed leptons $e_i$ those that are massless in the limit of zero Higgs VEV. The tree-level diagrams generating the two operators in \cref{eq:opsModelI} are shown in \cref{fig:modelI}.

The key challenge for this, or any other UV completion, lies in the fact that generic flavor structures of new Yukawa interactions typically predict low-multiplicity decays with much larger branching ratios. For example, replacing $e_1$ with $e_2$ in the second diagram of \cref{fig:modelI} would induce $\mu \to e \cS \to e (e^+ e^-)$. Thus, the observation of inverted branching-ratio hierarchies would pose a flavor puzzle~\cite{Altmannshofer:2024hmr}, indicating a specific underlying flavor symmetry and its associated symmetry-breaking pattern.

The SMEFT${}_X$ interactions in \cref{eq:opsModelI} display a distinctive flavor structure: the $\cS^3$ term links two lepton generations, while the $\cS$ term couples only to the first. We seek a configuration where $\mu \to e \,3\cS \to 7e$ dominates, with $\mu \to e\,2\cS \to 5e$ and $\mu \to e\,\cS \to 3e$ parametrically suppressed. Such pattern is enforced by a global $U(1)_\ell^3\equiv U(1)_e\times U(1)_\mu\times U(1)_\tau$ symmetry that is broken by two spurions, $\varepsilon_e$ and $\varepsilon_\mu$. The fields that carry nonzero charges under each of the global $U(1)$'s are 
\begin{align}
L_e:~~& [\ell_1]_{L_e}=[e_1]_{L_e}=[E]_{L_e}=1\,,
\\
L_\mu:~~ & [\ell_2]_{L_\mu}=[e_2]_{L_\mu}=[\phi]_{L_\mu}= 1\,, \quad [\cS]_{L_\mu}=\frac{1}{3}\,,
\\
L_\tau:~~& [\ell_3]_{L_\tau}=[e_3]_{L_\tau}=1\,.
\end{align}
In the SM, the flavor diagonal charged lepton Yukawa interactions are allowed by the $U(1)_\ell^3$ global symmetry, with the off-diagonal terms forbidden, thus defining the mass basis for charged leptons.

The $U(1)_\ell^3$ global symmetry is not exact, but rather explicitly broken by two small spurions, with charges $[ \varepsilon_\mu ]_{L_\mu} = -\frac{1}{3} $ and $[ \varepsilon_e ]_{L_e}=-1$. The two Wilson coefficients in \cref{eq:opsModelI} are therefore of order
\beq
\cC_{\cS}^{(3)}\sim {\mathcal O}(\varepsilon_e)\,, 
\qquad 
\cC_{e\cS}\sim {\mathcal O}(\varepsilon_\mu)\,.
\eeq
That is, the $S\to 2e$ decay rate is suppressed by $\varepsilon_\mu^2$, while the $\mu \to e  3\cS$ decay rate is suppressed by $\varepsilon_e^2$.
In contrast, the operators that would give rise to $\mu\to e \cS$ and $\mu \to e  2\cS$ decays are further suppressed, and are of order,
\beq
\label{eq:suppressed:expected}
\bar \ell_1 H e_2 \cS^\dagger\sim {\mathcal O}(\varepsilon_e \varepsilon_\mu^2)\,, 
\qquad 
\bar \ell_1 H e_2 \cS^{\dagger 2}\sim {\mathcal O}(\varepsilon_e \varepsilon_\mu)\,.
\eeq
If $\varepsilon_\mu \ll m_\mu/(4\pi \Lambda)$, there is a hierarchy among the exotic decays, $\Gamma(\mu \to e \cS)\ll \Gamma(\mu \to e  2\cS) \ll \Gamma(\mu \to e 3\cS)$, so that the $\mu\to 7e$ signature is the dominant one. Note that the $\varepsilon_e$ suppression does not enter in the relative ratios of these decay widths, and thus $\varepsilon_e$ can be large. 

In the UV completion with $\phi$ and $E$, the $U(1)_\ell^3$-invariant renormalizable interactions are
\begin{equation}
- \mathcal L \supset  y_{EH}\,\bar\ell_{1} H E +\lambda_{\phi \cS}\,\phi^\dagger \cS^3
+ \hermc\,,
\end{equation}
while at the linear order in spurions
\begin{equation}
-\mathcal{L} \supset \varepsilon_\mu \, y_{e\cS} \,\bar e_{1} \cS E + \varepsilon_e y_{E\phi}\, \bar e_{2}\,\phi\,E\,+\hermc\,.
\end{equation}
The $\phi$ and $E$ fields are assumed to be heavy and, after they are integrated out, generate at tree level the SMEFT${}_X$ operators in \cref{eq:opsModelI} with
\begin{equation}
\label{eq:cC7e:EFT}
\frac{\cC_{\cS}^{(3)}}{\Lambda^3} = \varepsilon_e \frac{y_{EH}\, y_{E\phi}^{\ast}\,\lambda^\ast_{\phi \cS}}{m_E\,m_\phi^2}\,,
\quad
\frac{\cC_{e\cS}}{\Lambda}= \varepsilon_\mu \frac{y_{EH}\, y_{e\cS}^\ast}{m_E}\,.
\end{equation}
We also assume that $3 m_\cS < m_\mu-m_e$,
 so that $\mu\to e 3\cS$ is kinematically allowed (the  hierarchy $m_\cS \ll m_{h,\phi, E}$ is set by hand).
Note that the  $\phi^\dagger \cS^3$ interaction is crucial for generating three on-shell $\cS$'s in the $\mu \to e 3\cS$ decay. We also assume that $\cS$ does not obtain a VEV, which can be easily arranged by the appropriate form of the $U(1)_\ell^3$  invariant scalar potential. If this is not the case, then $\mu\to e \cS$ and $\mu \to e  2\cS$ decays would also be possible at leading order in spurion insertions in the decay amplitudes, with the rates proportional to $\langle \cS\rangle^4$ and $\langle \cS\rangle^2$, respectively. 

In the limit of vanishing $\cS$ VEV, the $\mu \to e\cS$ only arises at higher order in spurion insertions. At quadratic order, we have 
\begin{equation}
\label{eq:ymuS}
-\mathcal{L} \supset  \varepsilon^*_e \varepsilon_\mu^2 \, y_{\mu \cS} \,\bar E e_{2} \cS^\dagger+\hermc\,, 
\end{equation}
which then gives rise to 
\beq
\mathcal{L} \supset  \frac{\cC_{\cS}^{(1)}}{\Lambda}\bar\ell_{1} H e_{2} \cS^\dagger+ \hermc\,,
\eeq
with
\beq
\frac{\cC_{ \cS}^{(1)}}{\Lambda}= \varepsilon^*_e \varepsilon^2_\mu \frac{y_{EH}^\ast \, y_{\mu \cS}}{m_E}\,,
\eeq
and thus to a suppressed $\mu \to e\cS$ rate, in accordance with \cref{eq:suppressed:expected}.

Note that the above flavor symmetry structure ensures radiatively stable hierarchies among different couplings, without fine-tuning. That is, even if one were to keep nonzero only the couplings linear in $\varepsilon_{e,\mu}$, namely $y_{e \cS}$ and $y_{E\phi}$,  the $y_{\mu \cS}$ would be generated at two-loops from a diagram involving  $\cS, \phi, E$ fields, including the $\varepsilon_e \varepsilon_\mu^2$ spurion insertion. The $\mu\to e \cS$, while suppressed, will thus never be completely zero (barring fine-tuned cancellations).

Since some of the couplings in the above model are highly suppressed, it is essential to verify whether the phenomenology resembles that of prompt $\mu \to 7e$ decays or if it necessarily involves displaced vertices. 
For concreteness, let us adopt as benchmark values $m_S = 30\,$MeV, $y_{EH} = y_{E\phi} = \lambda_{\phi \cS} = 1$, while requiring
\begin{align}
\label{eq:espmu:bench}
     \varepsilon_\mu \frac{v_\text{ew}}{m_E}  & \sim10^{\eminus5}\,,
     \\
     \label{eq:epse:epsmu2:ineq}
    \varepsilon_e \varepsilon^2_\mu \frac{v_\text{ew}}{m_E} y_{\mu \cS} & \lesssim 10^{\eminus14}\,.
\end{align}
 This benchmark yields $\cS\to e e$ that decays within the detector volume, with the $\cS$ lifetime given by
\begin{equation}
    c\tau_\cS =\frac{0.54\,\mathrm{mm}}{|y_{EH}y_{e\cS}^*|^2}
    \bigg( \frac{10^{\eminus4}}{\varepsilon_\mu} \bigg)^2
    \bigg( \frac{m_E}{\tev}\bigg)^2
    \bigg( \frac{30\,\mev}{m_\cS} \bigg)\,.
\end{equation}
The benchmark also satisfies the present beam-dump bounds, while the requirement in \cref{eq:epse:epsmu2:ineq} ensures that the current limits on $\mu\to 3e$ are satisfied, as can be seen from the results shown in Fig.~4b in Ref.~\cite{Knapen:2024fvh}.

For $y_{\mu \cS} \sim 1$, the inequality \cref{eq:epse:epsmu2:ineq} together with \cref{eq:espmu:bench} implies $\varepsilon_e \varepsilon_\mu \lesssim 10^{\eminus9}$. If we instead take the coupling in \cref{eq:ymuS} to be of a typical two-loop size, namely $y_{\mu \cS} \sim 1/(16\pi^2)^2$, then $\varepsilon_e \varepsilon_\mu \lesssim 10^{\eminus5}$. Constraints from the $Z$-pole observables require $v_\text{ew}/m_E \lesssim 10^{\eminus1}$~\cite{Greljo:2023adz, Greljo:2023bdy}, which, together with \cref{eq:espmu:bench}, implies $\varepsilon_\mu \gtrsim 10^{\eminus4}$. We therefore have $\varepsilon_e \lesssim 10^{\eminus5}$ for $y_{\mu \cS} \sim 1$, and $\varepsilon_e \lesssim 10^{\eminus1}$ for $y_{\mu \cS}$ that is two-loop suppressed, which we will use as another numerical benchmark.

With these benchmark values of parameters in hand, we can now translate the reach on the effective scale $\Lambda$ derived at the beginning of this section (see \cref{fig:7e_acceptance}) to a reach on $\phi$ and $E$ masses. From \cref{eq:cC7e:EFT}, setting $y_{EH} = y_{E\phi} = \lambda_{\phi \cS} = 1$ and $\cC_\cS^{(3)}=y_\mu$, the results in \cref{fig:7e_acceptance} imply
\begin{equation}
    \frac{y_\mu}{(1 \,\tev)^3}\simeq \frac{1}{(10\,\tev)^3} \gtrsim \varepsilon_e \frac{1}{m_E\,m_\phi^2}\,. 
\end{equation}
One thus obtains an observable $\mu\to 7e$ rate, for instance, for $m_E\sim 1\,\tev$ and $m_\phi \lesssim 100\,\gev$ ($\lesssim 10\,\tev$), if $\varepsilon_e \sim 10^{\eminus5}$($10^{\eminus1}$). The $\mu \to 7e$ searches can thus be complementary with direct searches for vector-like leptons (VLL) at the LHC~\cite{CMS:2024bni}. Note that when $m_\phi < m_E$, the LHC searches can be performed, besides the standard VLL signatures, also for the case where $E$ decays to a multi-lepton final state: $E \to \mu 3\cS$, with $\cS \to e e$ possibly displaced.

\subsection{Model II: Dark Photon with Flavor Violation}
\label{sec:DarkPhoton}
As the next example model, let us consider the dark sector, which contains a $U(1)_d$ gauge group. We focus on the minimal model, such that the dark sector consists just of a complex scalar $\Phi$, carrying a charge $[\Phi]=+1$ under $U(1)_d$. The $U(1)_d$ is spontaneously broken by a VEV  of the scalar field $v_\Phi$. The spectrum of states in the dark sector thus consists of a dark gauge boson (dark photon) $V_\mu$ with mass
\begin{equation}
    m_V = g_V v_\Phi \,,
\end{equation}
where $g_V$ is the $U(1)_d$ gauge coupling, and the dark Higgs $\varphi$ defined via $\Phi = (v_\Phi + \varphi)/\sqrt{2}$. With the scalar potential defined as
\begin{equation}
    V(\Phi)=\lambda_\Phi\lzm \Phi^\dagger\Phi - \frac{v_\Phi^2}{2}\dzm^2\,,
\end{equation}
the dark Higgs mass becomes
\begin{equation}
    m_\varphi = \sqrt{2 \lambda_\Phi} \, v_\Phi \,.
\end{equation}
The dark Higgs mass $m_\varphi$ can thus be small, i.e., $m_\varphi\ll v_\Phi$, if the scalar quartic coupling is small, $\lambda_\Phi\ll 1$.  

The dark sector is assumed to have two types of interactions with the visible sector. The flavor diagonal portal is due to 
a renormalizable kinetic mixing between dark photon and hypercharge field strengths, ${\cal L}\supset -\varepsilon V^{\mu\nu} B_{\mu\nu}/2c_W$, with $c_W$ the cosine of the Weinberg angle. This induces flavor-universal interactions inherited from the visible photon, enabling prompt $V \to e^+ e^-$ decays. 
 
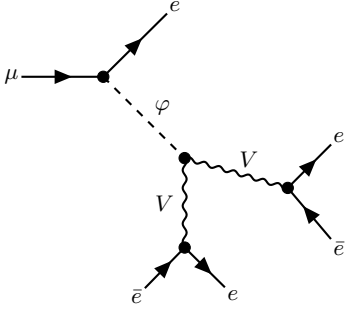
\begin{figure}[t]
\centering
	\begin{tikzpicture}[baseline=(a1.base), scale=0.34, transform shape]
			\begin{feynman}
			\vertex (a1);
			\vertex [left = 3.2cm of a1](a2) {\(\scalebox{2.8}{$\mu $}\)};
			\vertex [above right = 3.5cm of a1](a3) {\(\scalebox{2.8}{$e $}\)};
			\vertex [below right = 4.5cm of a1](a4);
			\vertex [below  = 3.5cm of a4](a5);
			\vertex [above right = 4.5cm of a5](a6);
			\vertex [below right = 1.2cm of a6](a7);
			\vertex [below right = 2.2cm of a5](a6BR) {\(\scalebox{2.8}{$e $}\)};
			\vertex [below left = 2.2cm of a5](a6BL) {\(\scalebox{2.8}{$\bar e $}\)};
			\vertex [right = 2.0cm of a7](a7BL);
			\vertex [above = 1.7cm of a7BL](a7BLabove) {\(\scalebox{2.8}{$e$}\)};
			\vertex [below = 2.0cm of a7BL](a7BLbelow) {\(\scalebox{2.8}{$\bar e $}\)};
			\diagram*{
			(a2) -- [fermion, thick] (a1)--[fermion, thick](a3),
			(a1)--[dashed, thick, edge label=\(\scalebox{2.8}{$~\varphi$}\)](a4),
			(a5)--[photon, thick, edge label=\(\scalebox{2.8}{$V~$}\)](a4),
			(a4)--[photon, thick, edge label=\(\scalebox{2.8}{$V_{\phantom{aa}}$}\)](a7),
			(a5)--[fermion, thick](a6BR),
			(a6BL)--[fermion, thick](a5),
			(a7)--[fermion, thick](a7BLabove),
			(a7BLbelow)--[fermion, thick](a7)
			};
			\draw[fill=black] (a1) circle(2.2mm);
			\draw[fill=black] (a4) circle(2.2mm);
			\draw[fill=black] (a7) circle(2.2mm);
			\draw[fill=black] (a5) circle(2.2mm);
			\end{feynman}
	\end{tikzpicture}
    \caption{Feynman diagrams corresponding to the cascade transition given by \cref{eq:DM_cascade_1}. See \cref{sec:DarkPhoton} for details.}
    \label{fig:Dark_photon_model}
\end{figure}

Additionally, there is a flavor-violating portal between dark and visible sectors, due to a higher-dimensional gauge-invariant operator, suppressed by a heavy scale $\Lambda$,
\begin{equation}
\label{eq:cLFV:portal}
    \mathcal{L}_{\sscript{port.}} \supset \frac{\cC_{e\mu}}{\Lambda^2} \, (\bar \ell_1 H e_2 )\, \Phi^\dagger \Phi + \hermc\,.
\end{equation}
Once $\Phi$ and $H$ acquire VEVs, this operator induces cLFV interactions, coupling SM leptons with the dark Higgs, 
\begin{equation}
  \mathcal{L}_{\sscript{port.}} \supset \frac{\cC_{e\mu}}{\Lambda^2} \frac{v_\text{ew} v_\Phi}{\sqrt2} \, (\bar e_L \mu_R)\, \varphi+\hermc\,.
\end{equation}
For the mass hierarchy
\begin{equation}
\label{eq:mass:ranges:dark:photon}
    m_\mu  -m_e > m_\varphi > 2 m_V > 4 m_e \,,
\end{equation}
the dominant muon decay involving dark sector states is via a cascade (see~\cref{fig:Dark_photon_model})
\begin{equation}\label{eq:DM_cascade_1}
    \mu \to \varphi\, e \,\to\,  (VV) \, e  \,\to\, (e^+ e^-) \, (e^+ e^-) \, e \,,
\end{equation}
with the intermediate states $\varphi$ and $V$ produced on-shell. 

The mass hierarchy in \cref{eq:mass:ranges:dark:photon} is naturally realized for $v_\Phi \sim m_\mu$ and moderate couplings, $g_V, \lambda_\Phi \sim \mathcal{O}(0.3)$, provided the inequality $\lambda_\Phi > g^2_V$ holds. The last two steps of the cascade in \cref{eq:DM_cascade_1} proceed with $\sim 100\%$ branching fractions. The overall $\mu\to 5e$ decay rate is thus controlled by the first decay in the cascade, $\mu \to  e \varphi$. Assuming that this has a branching ratio $\mathrm{BR}(\mu \to e \varphi)=\mathcal{O}(10^{\eminus15})$, while all the subsequent decays occur with unit probability and are prompt, this setup probes scales of  order
\begin{equation}
    \Lambda \gtrsim \sqrt{10^{11}\,\mathrm{TeV} \times v_\Phi} \, ,
\end{equation}
where we have taken $\cC_{e\mu}\simeq y_\mu$. 

The above model represents a variation of the model introduced in Ref.~\cite{Hostert:2023gpk}.\footnote{The only difference is related to the interpretation of the effective scale, which in  Ref.~\cite{Hostert:2023gpk} was taken to be $\Lambda_\sscript{eff}= \Lambda^2/v_\Phi$.} Additionally, Ref.~\cite{Hostert:2023gpk} also performed a detailed phenomenological analysis of $\mu\to 5e$ signal reach at Mu3e, with $\mathrm{BR}(\mu \to e V)=\mathcal{O}(10^{\eminus12})$ taken as a conservative estimate if all accidental backgrounds are accepted, while with kinematic cuts $\mathrm{BR}(\mu \to e V)=\mathcal{O}(10^{\eminus15})$ may well be within reach.  Variations of the above general setup are possible: an intriguing possibility is that the SM fermions are charged under $U(1)_d$. In that case, it is possible for the cLFV portal to the dark sector to not be quadratic in $\Phi$, as in \cref{eq:cLFV:portal}, but rather linear in $\Phi$. In Appendix \ref{sec:AppU(1)mu-e} we provide further details for the case where the dark gauge group is identified with $U(1)_{\mu-e}$.

\subsection{Model III: Thermal Inelastic Dark Matter}
\label{sec:model_dark_higgs}

An elegant dark matter (DM) extension can be incorporated into the Model II presented above. Following Ref.~\cite{Duerr:2019dmv}, we introduce a SM-singlet vector-like fermion $\psi_{L,R}$ carrying charge $+1$ under the dark gauge symmetry $U(1)_d$. If the scalar $\Phi$ carries a $U(1)_d$ charge $[\Phi]=+2$, the dark sector Lagrangian contains Yukawa-type interactions between $\Phi$ and dark fermions:
\begin{equation}
    -\mathcal{L} \supset 
     m_D \, \bar\psi_L \psi_R 
    + \lambda_L \Phi \, \bar\psi_L \psi_L^c
    + \lambda_R \Phi \, \bar\psi_R \psi_R^c
    + \hermc \, .
\end{equation}
Imposing an exchange symmetry $\psi_L \leftrightarrow \psi_R$, enforces $\lambda_\psi \equiv \lambda_L = \lambda_R$ and leads to maximal mixing between $\psi_{L,R}$, i.e., with mixing angle $\theta = \pi/4$. The resulting mass eigenstates are Majorana fermions $\chi_{1,2}$ with masses
\begin{equation}
    m_{\chi_{2,1}} = \lambda_\psi v_\Phi \pm m_D \,.
\end{equation}
The lighter state $\chi_1$ is absolutely stable due to a residual $Z_2^\psi$ symmetry.  

Diagonalizing the gauge interaction yields \textit{solely} off-diagonal couplings with the dark photon
\begin{equation}
 \mathcal{L} \supset g_V \bar\chi_1 \gamma^\mu \chi_2 \, V_\mu \,,
\end{equation}
which presents a key feature for realizing a model of light thermal DM. In conventional light dark matter scenarios, the relic abundance is typically set by the elastic freeze-out through annihilations such as $\chi\chi \to e^+e^-$. However, such processes are strongly constrained by CMB observations, since late-time annihilations inject energy during recombination. This problem is avoided in the above model: the relic abundance is instead determined by efficient $\chi_1$--$\chi_2$ coannihilation, which freezes out in the early universe. After freeze-out, the heavier $\chi_2$ states quickly decay away, leaving only the stable $\chi_1$ component and no significant annihilation in the late universe. This mechanism reconciles the correct relic abundance with the indirect detection limits. Moreover, direct-detection signals are strongly suppressed due to the inelastic nature of the scattering.  For other realizations of light thermal DM, see Refs.~\cite{Tucker-Smith:2001myb, Hochberg:2014kqa, Izaguirre:2015zva, Davighi:2024zip, Davighi:2025awm,Lu:2023cet}.

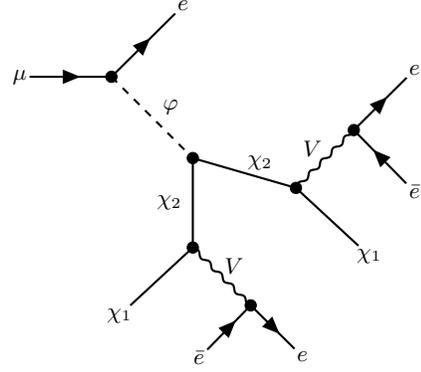
\begin{figure}[t]
\centering
	\begin{tikzpicture}[baseline=(a1.base), scale=0.34, transform shape]
			\begin{feynman}
			\vertex (a1);
			\vertex [right = 0.0cm of a1](a1EFT);
			\vertex [left = 3.2cm of a1EFT](a2EFT) {\(\scalebox{2.8}{$\mu $}\)};
			\vertex [above right = 3.5cm of a1EFT](a3EFT) {\(\scalebox{2.8}{$e $}\)};
			\vertex [below right = 4.5cm of a1EFT](a4EFT);
			\vertex [below  = 3.5cm of a4EFT](a5EFT);
			\vertex [above right = 4.5cm of a5EFT](a6EFT);
			\vertex [below right = 1.2cm of a6EFT](a7EFT);
			\vertex [below right = 3.2cm of a5EFT](a6BREFT);
			\vertex [below left = 3.2cm of a5EFT](a6BLEFT) {\(\scalebox{2.8}{$\chi_1 $}\)};
			\vertex [right = 2.0cm of a7EFT](a7BLEFT);
			\vertex [above = 1.7cm of a7BLEFT](a7BLaboveEFT);
			\vertex [below = 2.0cm of a7BLEFT](a7BLbelowEFT);
			\vertex [below right = 2.4cm of a6BREFT](a8BREF) {\(\scalebox{2.8}{$e$}\)};
			\vertex [below left = 2.4cm of a6BREFT](a8BLEF) {\(\scalebox{2.8}{$\bar e$}\)};
			\vertex [below right = 3.2cm of a7EFT](a10EFT) {\(\scalebox{2.8}{$\chi_1 $}\)};
			\vertex [above right = 3.2cm of a7EFT](a11EFT);
			\vertex [above right = 2.9cm of a11EFT](a12ar) {\(\scalebox{2.8}{$e$}\)};
			\vertex [below right = 2.9cm of a11EFT](a12br) {\(\scalebox{2.8}{$\bar e$}\)};
			\diagram*{
			(a2EFT)-- [fermion, thick] (a1EFT)--[fermion, thick](a3EFT),
			(a1EFT)--[dashed, thick, edge label=\(\scalebox{2.8}{$~\varphi$}\)](a4EFT),
			(a5EFT)--[plain, thick, edge label=\(\scalebox{2.8}{$\chi_2~$}\)](a4EFT),
			(a4EFT)--[plain, thick, edge label=\(\scalebox{2.8}{$\chi_2$}\)](a7EFT),
			(a5EFT)--[photon, thick, , edge label=\(\scalebox{2.8}{$V$}\)](a6BREFT),
			(a5EFT)--[plain, thick](a6BLEFT),
			(a6BREFT)--[fermion, thick](a8BREF),
			(a8BLEF)--[fermion, thick](a6BREFT),
			(a7EFT)--[plain, thick](a10EFT),
			(a7EFT)--[photon, thick, edge label=\(\scalebox{2.8}{$V$}\)](a11EFT),
			(a11EFT)--[fermion, thick](a12ar),
			(a12br)--[fermion, thick](a11EFT)
			};
			\draw[fill=black] (a1EFT) circle(2.2mm);
			\draw[fill=black] (a4EFT) circle(2.2mm);
			\draw[fill=black] (a5EFT) circle(2.2mm);
			\draw[fill=black] (a6BREFT) circle(2.2mm);
			\draw[fill=black] (a7EFT) circle(2.2mm);
			\draw[fill=black] (a11EFT) circle(2.2mm);
			\end{feynman}
	\end{tikzpicture}
    \caption{Feynman diagram giving the cascade decay in \cref{eq:DM_cascade_2} for the thermal inelastic dark matter model. See \cref{sec:model_dark_higgs} for details.}
    \label{fig:DM_model_diagrams}
\end{figure}

A possible mass hierarchy
\begin{equation}\label{eq:1To2}
\small
    m_\mu - m_e > m_\varphi> 2 m_{\chi_2} > 2(m_V + m_{\chi_1})\,\, \text{and}\,
\, m_V > 2m_e
\end{equation}
allows for exotic $1 \to 2 $ cascade decays. In this regime, the exotic muon decay proceeds via the sequence (see \cref{fig:DM_model_diagrams})
\begin{equation}\label{eq:DM_cascade_2}
    \begin{alignedat}{2}
    \mu \to \varphi \, e 
     \to  (\chi_2 \chi_2) \, e 
     &\to  (\chi_1  V)(\chi_1 V) \, e 
     \\
     &\to  \chi_1(e^+ e^-) \, \chi_1(e^+ e^-) \, e \,.
    \end{alignedat}
\end{equation}
The first decay is induced via the cLFV operator in \cref{eq:cLFV:portal}. Since the $\chi_1$ states carry away missing energy, the resulting signature is experimentally more challenging. An important experimental handle is that the $e^+e^-$ invariant mass reproduces the dark photon mass, though such bump hunting requires more statistics than a zero-background search would. The decays $\varphi\to 2\chi_2$,  $\chi_2\to V\chi_1$, and $V\to 2e$ typically have large branching fractions and are prompt.

Alternatively, the mass spectrum may not follow \cref{eq:1To2}. For example, $m_V$ may be heavier such that the $\chi_2\to \chi_1 2e$ is then a 3-body decay due to a tree-level exchange of an off-shell dark photon $V$. For sufficiently heavy $V$, these decays will be displaced, providing an additional handle in experimental searches.  In addition, $\varphi$ may be heavier than $m_\mu$, in which case $\mu \to e \chi_2 \chi_2$ proceeds via contact interaction due to an off-shell $\varphi$; a scenario that would require a much lower cLFV scale in \cref{eq:cLFV:portal}. Finally, note that dark Higgs couples in a flavor-universal way to $\chi_{1,2}$, so that $\mu \to e+\mathrm{inv}$ decays are generated at a comparable level with the cascade in \cref{eq:DM_cascade_2}. However, this signature is less promising experimentally. 

\begin{figure}[t]
\centering
\includegraphics[width=0.495\textwidth]{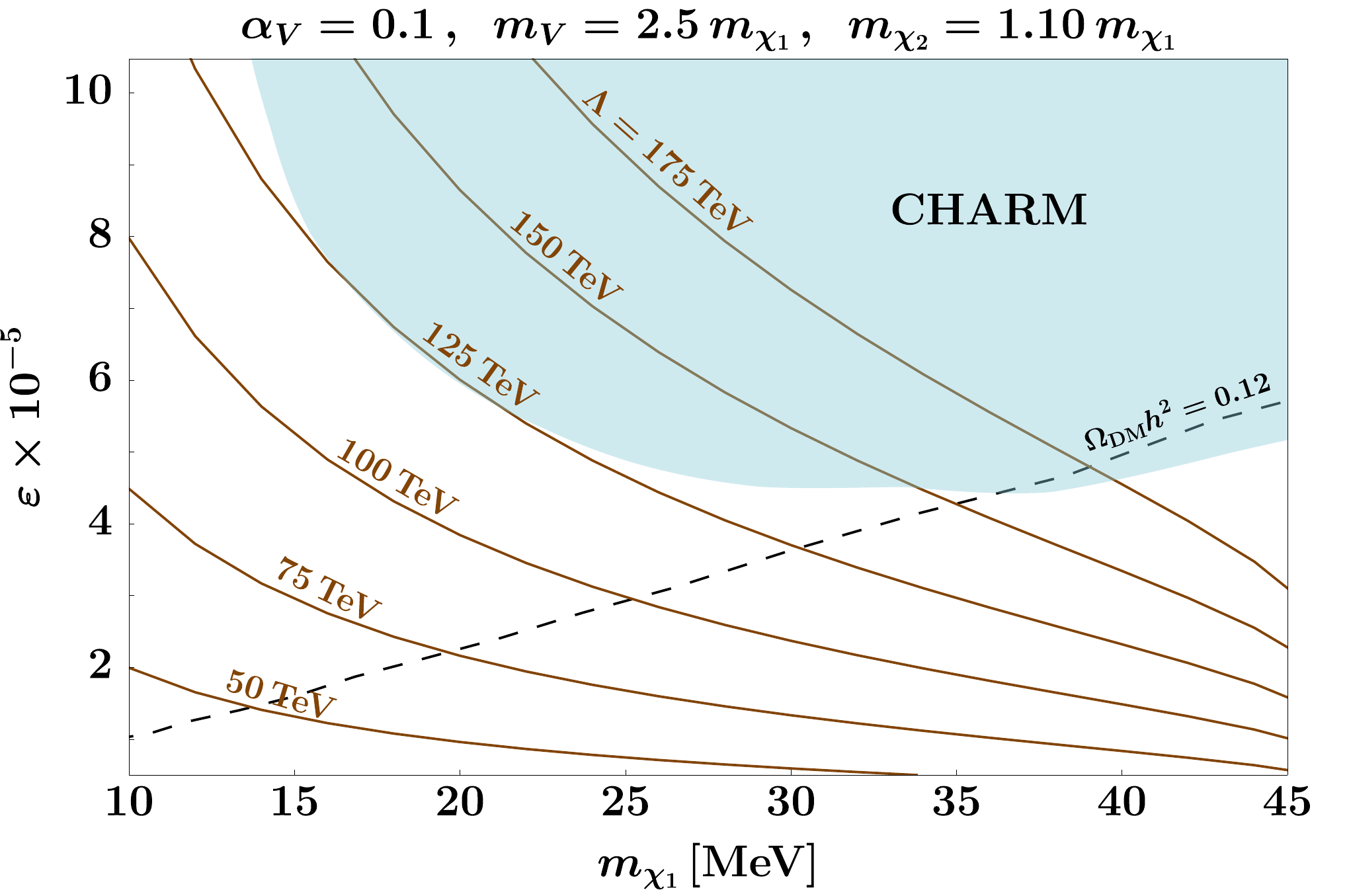} 
\caption{Exclusion contours in the $(m_{\chi_1}, \varepsilon)$ plane for the benchmark scenario defined in \cref{eq:DM_benchmark}. The shaded region denotes the constraints from CHARM~\cite{Duerr:2019dmv}, while the solid lines correspond to the projected sensitivity for different values of the cLFV scale $\Lambda$. The dashed line denotes the parameter space compatible with the observed dark matter relic abundance. 
}
\label{fig:DM_plot}
\end{figure}

The detailed DM phenomenology of the above model, including the viable parameter space that reproduces the observed relic abundance while satisfying all complementary constraints, is discussed in Ref.~\cite{Duerr:2019dmv}. To illustrate the relevance of the exotic muon decay, we take the benchmark shown in the top-right Fig.~6 in Ref.~\cite{Duerr:2019dmv}. In this benchmark, the values of the parameters are set to
\begin{equation}\label{eq:DM_benchmark}
g_V = \sqrt{\frac{2}{5}\pi}\,,
\quad
m_V = 2.5\,m_{\chi_1}\,,
\quad
m_{\chi_2} = 1.10\,m_{\chi_1}\,.
\end{equation}
In addition, we set $m_\varphi = 0.1\,\gev$, and plot in \cref{fig:DM_plot} exclusion contours in the $(m_{\chi_1},\varepsilon)$ plane for various values of $\Lambda$ defined in \cref{eq:cLFV:portal} for $\cC_{e\mu} = y_\mu$ (brown solid lines), requiring at least three electrons within the detector volume\footnote{In this scenario, only one of two $\chi_2$ produced in $\mu \to e 2\chi_2$ transition, decays inside the detector. We adopt $L_{\sscript{det}} = 0.7\,\mathrm{m}$ as a representative detector length. For this choice of parameters, $\mu \to 5e + \text{inv}$ is subleading due to the long $\chi_2$ lifetime $L_{\sscript{det}}/(c \tau_{\chi_2}) < 10^{\eminus4}$ for the parameter ranges shown in \cref{fig:DM_plot}.}
\begin{equation}\label{eq:DMbr3e}
\begin{split}
    \mathrm{BR}(\mu\to 3e+\text{inv})_\sscript{eff} =& \mathrm{BR}(\mu \to e \varphi) \times
    \\
    \times & \Big[1 - \exp\big(\eminus {L_{\sscript{det}}}/{ \beta \gamma  c \tau_{\chi_2}}\big) \Big]\,.
    \end{split}
\end{equation}
Here, $\tau_{\chi_2}$ denotes the lifetime of $\chi_2$ decaying through an off-shell $V$~\cite{Izaguirre:2015zva}
\begin{equation}
    c\tau_{\chi_2} \simeq \frac{15\pi}{4\varepsilon^2}\frac{1}{\alpha_{\sscript{em}}\alpha_V}\frac{m_V^4}{\Delta^5}\,,
    \qquad
    \Delta=m_{\chi_2}-m_{\chi_1}\,,
\end{equation}
while the $\mu\to e \varphi$ branching ratio is~\cite{Hostert:2023gpk}
\begin{equation}
    \mathrm{BR}(\mu\to e\varphi)\simeq 24\pi^2 \frac{ v_\text{ew}^4}{\Lambda^4}\frac{v_\Phi^2}{m_\mu^2}\lzm 1-\frac{m_\varphi^2}{m_\mu^2} \dzm^2\,.
\end{equation}
In the numerical analysis we set $v_\Phi=m_\varphi$.

The dashed black line in \cref{fig:DM_plot} gives the contour where the correct relic abundance is obtained, while the blue shaded region is excluded by CHARM constraints~\cite{Tsai:2019buq}. As illustrated in the plot, for $\Lambda \lesssim 150~\text{TeV}$, the exotic muon decay modes constitute the most powerful probes. Our simple estimate of the scale $\Lambda$, indicated by the solid brown contours in \cref{fig:DM_plot}, is obtained under the assumption of \cref{eq:DMbr3e} set to $10^{\eminus15}$. The experimental search in this regime is challenging due to the relatively low $p_T$ electrons from a displaced vertex, which requires a dedicated sensitivity study with a careful evaluation of SM backgrounds. As already anticipated, we explicitly checked that the $\mu \to e + \text{inv}$ channel is not competitive within the parameter space of interest.

Finally, a comment is in order regarding the $\mu \to 5e + \text{inv}$ channel. Although this signature is experimentally cleaner, its branching ratio becomes comparable to that of $\mu \to 3e + \text{inv}$ only when $L_{\sscript{det}} \sim \beta\gamma (c\tau_{\chi_2})$. This occurs in regions of parameter space with larger mass splittings $\Delta$, which are, however, constrained by proton on target beam-dump experiments (cf. Fig.~6 of Ref.~\cite{Duerr:2019dmv} and Ref.~\cite{Tsai:2019buq}). Variations of the model that relax these bounds could involve modified $U(1)$ charge assignments for quarks, leptons, and $\chi$ fields, a possibility we leave for future investigation.

\subsection{Model IV: On- and Off-Shell ALP}
\label{sec:model_off_shell_ALP}

The decay process $\mu \to e a$, with $a$ being a light pNGB,\footnote{Here, we use the more common notation, $a$, for pNGB, while in \cref{sec:EFT:lightNP} we denoted it with $\cP$.} is a well-studied example of an exotic muon decay that has received considerable attention in the literature \cite{Calibbi:2020jvd, Jho:2022snj, Xing:2022rob, Panci:2022wlc, Hill:2023dym, Knapen:2023zgi, Knapen:2024fvh, Bigaran:2025uzn, MartinCamalich:2025srw, Bonnefoy:2019lsn} (for tau decays involving ALPs, see, e.g.,~\cite{Belle:2025bpu, Ema:2025bww}).
Examples include $a$ that is an LFV QCD axion, an axiflavon, a leptonic familon, or a majoron~\cite{Calibbi:2020jvd}. In all of these examples, $a$ is long-lived and escapes the detector, resulting in a $\mu \to e+$inv decay signature, where the electron is mono-energetic with energy $E_e\simeq m_\mu/2$. 

Here, we are interested instead in the possibility that $a$ decays within the detector volume via $a\to \gamma\gamma$. While such models are already part of the experimental search program, e.g., at MEG II~\cite{MEG:2020zxk}, they do face significant theoretical and phenomenological challenges. The parameter region compatible with the prompt $a\to\gamma\gamma$ decays typically requires a symmetry-breaking scale of order $v_\Phi\sim 10\,\gev$, which is challenging to realize in UV-complete setups. For $v_\Phi \gtrsim 1\,\tev$, the ALP instead becomes long-lived, leading to displaced signatures that are already excluded by beam-dump and supernova constraints~\cite{Heeck:2017xmg}. 
 
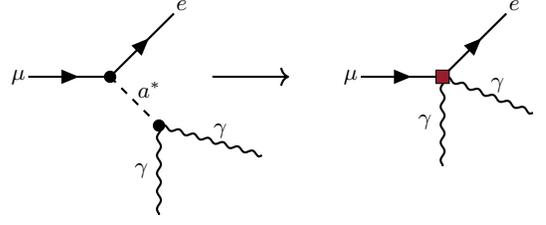
\begin{figure}[t]
    \centering
	\begin{tikzpicture}[baseline=(a1.base), scale=0.34, transform shape]
			\begin{feynman}
			\vertex (a1);
			\vertex [left = 3.2cm of a1](a2) {\(\scalebox{2.8}{$\mu $}\)};
			\vertex [above right = 3.5cm of a1](a3) {\(\scalebox{2.8}{$e $}\)};
			\vertex [below right = 2.7cm of a1](a4);
			\vertex [below  = 3.5cm of a4](a5);
			\vertex [above right = 4.5cm of a5](a6);
			\vertex [below right = 1.2cm of a6](a7);
			\vertex [right = 13.0cm of a1](a1EFT);
			\vertex [left = 3.2cm of a1EFT](a2EFT) {\(\scalebox{2.8}{$\mu $}\)};
			\vertex [above right = 3.5cm of a1EFT](a3EFT) {\(\scalebox{2.8}{$e $}\)};
			\vertex [below = 3.5cm of a1EFT](a4EFT);
			\vertex [below right = 3.5cm of a1EFT](a5EFT) ;
			\vertex [above right = 1.5cm of a5EFT](a6EFT);
			\vertex [right = 4.0cm of a1](a1arr);
			\vertex [right = 3.0cm of a1arr](a1arr2);
			\diagram*{
			(a2) -- [fermion, thick] (a1)--[fermion, thick](a3),
			(a1)--[dashed, thick, edge label=\(\scalebox{2.8}{$a^*$}\)](a4),
			(a5)--[photon, thick, edge label=\(\scalebox{2.8}{$\gamma~$}\)](a4),
			(a4)--[photon, thick, edge label=\(\scalebox{2.8}{$\gamma~$}\)](a7),
			(a2EFT)-- [fermion, thick] (a1EFT)--[fermion, thick](a3EFT),
			(a4EFT)--[photon, thick, edge label=\(\scalebox{2.8}{$\gamma~$}\)](a1EFT),
			(a1EFT)--[photon, thick, edge label=\(\scalebox{2.8}{$\gamma~$}\)](a6EFT)
			};
			\draw[fill=black] (a1) circle(2.2mm);
			\draw[fill=black] (a4) circle(2.2mm);
			\draw[fill=red!90!gray] (a1EFT) +(-2.6mm,-2.6mm) rectangle +(2.6mm,2.6mm);
			\draw[->, thick] (a1arr) -- (a1arr2);
			\end{feynman}
	\end{tikzpicture}
    \caption{\textit{Left:} Feynman diagram generating $\mu \to e\gamma\gamma$ decay via an off-shell ALP exchange. \textit{Right}: the corresponding local contact interaction obtained in the EFT limit. See \cref{sec:model_off_shell_ALP} for details.}
    \label{fig:off_shell_ALP_diagrams}
\end{figure}

Below, we extend the parameter space to include off-shell ALP production $\mu \to e a^*\to e 2\gamma$. For heavy enough ALP, the astrophysical and beam-dump constraints are avoided, so that this can be viewed as a benchmark model for three body $\mu\to e 2\gamma$ exotic decays.\footnote{Incidentally, this is also a realization of exotic muon decays in the intermediate EFT regime mentioned in \cref{sec:SMEFT}, since the ALP is a mediator that has a mass between the muon mass and the electroweak scale.} Note, however, that this benchmark will typically face stringent constraints from $\mu \to e \gamma$ in concrete UV models, as we show below. Furthermore, unlike in the case of an on-shell ALP production, now there is no narrow peak in the di-photon spectrum, so that the experimental search needs to take this into account.

In the rest of this section, we perform a phenomenological analysis of the $\mu \to e (a \to 2\gamma)$ decays, both in the on- and off-shell ALP production regimes. For concreteness, we work within a minimal UV-complete model for a leptophilic ALP, so that we have full control of both the size of the lepton-flavor violating couplings of the ALP, as well as of the ALP--photon couplings. 
  
In this model, the SM is extended by the $U(1)_\mu\times U(1)_\tau$ global symmetry, while the field content is extended by a complex scalar $\Phi$ and two vector-like leptons, $E$ and $\Psi$. The complex scalar has $U(1)_\mu\times U(1)_\tau$ charges $[\Phi]=(-1,0)$, so that $U(1)_\mu$ is spontaneously broken once $\Phi$ obtains a VEV, denoted by $v_\Phi$.\footnote{We use the short-hand notation $[\Phi]=([\Phi]_{L_\mu}, [\Phi]_{L_\tau})$.} The SM leptons are charged in the usual way under $U(1)_\mu\times U(1)_\tau$: the muon doublet and singlet have charges $[\ell_{2}]=[e_{2}]=(+1,0)$, while the third generation lepton fields have charges $[\ell_{3}]=[e_{3}]=(0,+1)$. Both $E$ and $\Psi$ have the same SM quantum numbers as the right-handed SM leptons, i.e., $E\sim(\bm1,\bm1)_{\eminus1}$ and $\Psi\sim(\bm1,\bm1)_{\eminus1}$. With regard to $U(1)_\mu\times U(1)_\tau$, $E$ is a singlet, while $\Psi$ has chiral charges given as $[\Psi_L]=(-1 + q,0)$ and $[\Psi_R]=(q,0)$, a characteristic feature of KSVZ-type models where the PQ symmetry acts chirally on the heavy fermions~\cite{Kim:1979if, Shifman:1979if}. We further assign a $\mathbb Z_2$ parity under which $\Psi_{L,R}$ are odd, while all the other fields are even. The $\mathbb Z_2$ is weakly broken, such that $\Psi$ decays into lighter SM states before the Big Bang Nucleosynthesis (BBN) through interactions of the form $-\cL_\Psi\supset \bar\ell_1 H \Psi_R$. This minor point does not significantly impact the rest of our discussion.

With these field assignments, the additional renormalizable interactions are
\begin{equation}
    \begin{alignedat}{2}
        -\cL &\supset y_{EH}\bar\ell_1 H E_R+y_{E\Phi}\bar E_L \Phi e_2+m_E\bar E_L E_R
        \\&~+y_\Phi\Phi\bar\Psi_L\Psi_R+\hermc\,,
    \end{alignedat}
\end{equation}
where $y_{EH}$, $y_{E\Phi}$ and $y_\Phi$ denote Yukawa-like couplings, while $m_E$ is the vector-like mass of $E$. Note that the $U(1)_\mu\times U(1)_\tau$ also enforces the SM Yukawas to be flavor diagonal, and thus $\ell_1$ and $e_2$ correspond to the SM mass eigenstates of the first and the second generation leptons, respectively.

Integrating out $E$ at tree level yields a dimension-five operator that links the scalar $\Phi$ to the SM leptons (the diagram is as in \cref{fig:modelI} (right), but replacing $\cS$ with $\Phi$),
\begin{equation}\label{eq:ALP_dim-5}
    \cL_{\sscript{eff}}\supset \frac{y_{EH}y_{E\Phi}}{m_E}(\bar\ell_1 H e_2)\Phi+\hermc\,.
\end{equation}
Once $H$ and $\Phi$ obtain VEVs, where
\begin{equation}
     \langle\Phi\rangle=\frac{v_\Phi+\varphi}{\sqrt2}e^{ia/v_\Phi}\,,
\end{equation}
the operator in \cref{eq:ALP_dim-5} induces ALP couplings to $\mu$ and $e$
\begin{equation}\label{eq:ALP_mu_e}
    \cL_{\sscript{ALP}}\supset \frac{\partial_\mu a}{v_\Phi}\bar e\gamma^\mu (g_{e\mu}^V+g_{e\mu}^A\gamma_5) \mu\,,
 \end{equation}
 where the two couplings are equal, and given by
 \beq
    g_{e\mu}^{V}=  g_{e\mu}^{A}=\frac{y_{EH}y_{E\Phi}}{4m_\mu}\frac{v_\text{ew}v_\Phi}{m_E}\,,
 \eeq
 so that the ALP couples only to the right-handed leptons.

The dimension-5 operator in \cref{eq:ALP_dim-5} also induces off-diagonal terms in the charged lepton mass matrix. To avoid large cancellations when diagonalizing the mass matrix, one requires
$\left|y_{EH}y_{E\Phi} v_\Phi/ m_E \right| \lesssim y_\mu$.\footnote{Note that this is a less stringent requirement than the more common $y_{\mu e} y_{e\mu}\lesssim y_e y_\mu$ \cite{Cheng:1987rs, Harnik:2012pb}. The reason is that in our case, the charged lepton mass matrix has only one off-diagonal entry that is nonzero.} For simplicity, we assume that $y_{EH}y_{E\Phi} v_\Phi/ m_E$ is even well below this naturalness limit, so that the charged lepton mass matrix can be diagonalized in the small mixing approximation. The misalignment between the interaction and the mass basis is then given by the left-handed mixing
\begin{equation}
    \theta_{e\mu}\approx-\frac{1}{\sqrt2}\frac{y_{EH}y_{E\Phi}}{y_\mu}\frac{v_\Phi}{m_E}\,.
\end{equation}
The right-handed mixing is further suppressed, and is given by $\theta_R=\theta_{e\mu} y_e/y_\mu$.
Since the ALP coupling to leptons is controlled by the right-handed mixing, with the ALP--electron coupling $\propto \theta_R$, the $a\to e^+e^-$ decays are suppressed and can be neglected in the region of parameter space that we are interested in.  

\begin{figure}[t]
\centering
\includegraphics[width=0.495\textwidth]{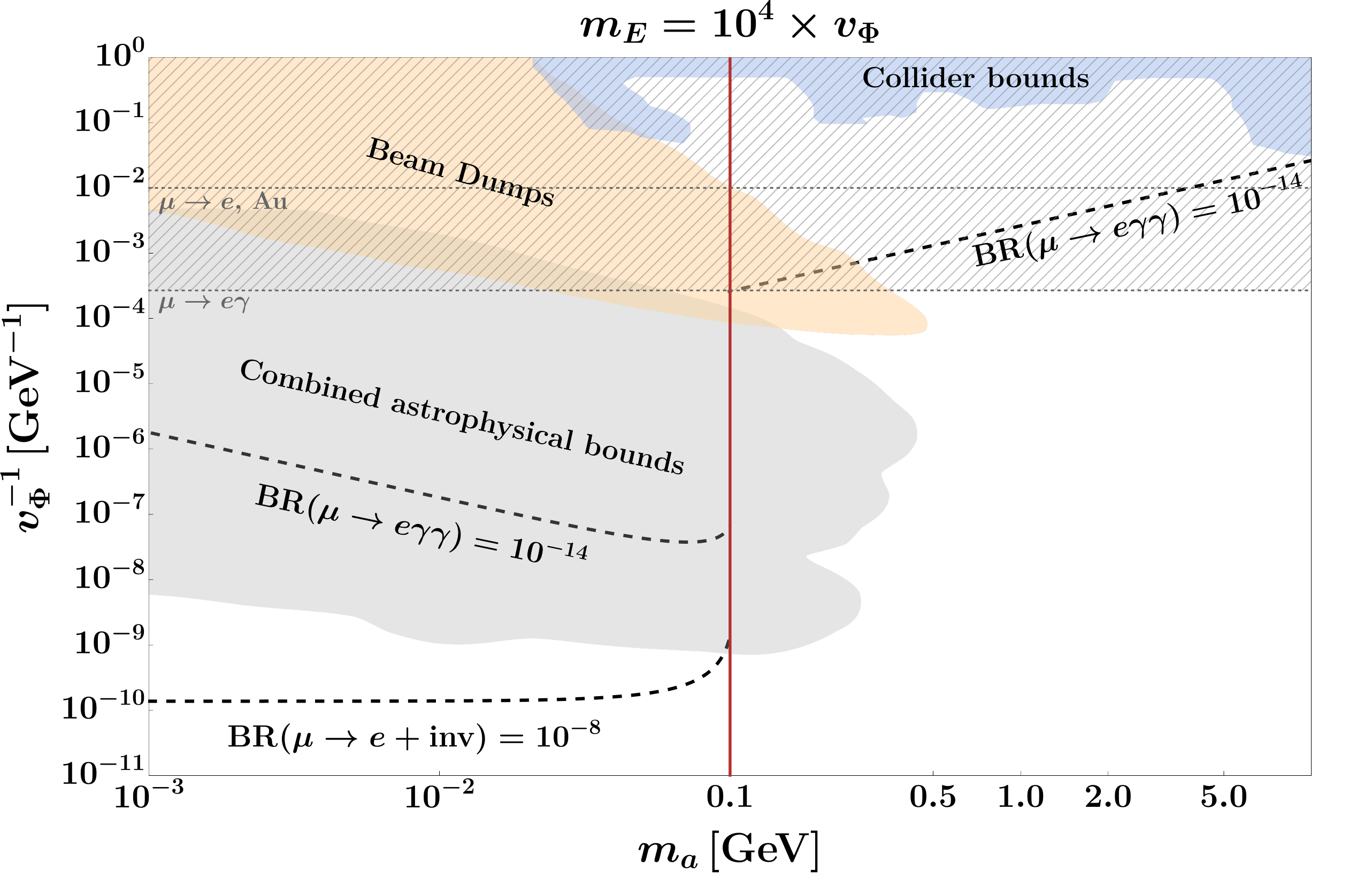} 
\caption{The reach in heavy scalar VEV, $v_\Phi$, as a function of ALP mass, $m_a$, for possible future bounds on $\mu \to e \gamma \gamma$ and $\mu \to e +\mathrm{inv}$ branching ratios (dashed lines), within a KSVZ-like axion model, taking as a benchmark $m_E = 10^4 \, v_\Phi$, see  \cref{sec:model_off_shell_ALP}. The constraints from astrophysics are shown as a gray shaded region, from beam dumps as an orange region, and from collider searches as a blue region~\cite{AxionLimits}. The indirect bounds from $\mu\to e$ conversion and from $\mu\to e\gamma$ constraints (hatched region) assume no cancellations against other UV contributions. 
}
\label{fig:alp_plots_combined}
\end{figure}

Finally, the triangle diagram with vector-like fermion $\Psi$ running in the loop generates the anomalous couplings of the ALP to the hypercharge gauge boson
\begin{equation}
    \cL_{\sscript{ALP}}\supset \frac{g_1^2}{16\pi^2}\frac{a}{v_\Phi}B_{\mu\nu}\widetilde B^{\mu\nu}\,,
\end{equation}
with $g_1$ denoting the $U(1)_Y$ gauge coupling. 
After EWSB, these induce couplings of ALP to photons given by
\begin{equation}
    \cL_{\sscript{ALP}}\supset   \frac{\alpha_\sscript{em}}{4\pi}  \frac{a}{v_\Phi}F_{\mu\nu}\widetilde F^{\mu\nu}\,,
\end{equation}
which in turn lead to $a^{(*)}\to\gamma\gamma$ decays, and enable ALP production in astrophysical systems, beam-dump experiments, and colliders. The corresponding constraints are shown as shaded regions in \cref{fig:alp_plots_combined}.

Next, we estimate the reach of rare muon experiments in the on-shell and off-shell ALP regimes. 

\vspace{0.2cm}
\noindent
{\textbf{Off-shell ALP.}} The branching ratio for $\mu \to e a^* \to e(\gamma\gamma)$ decay is given by
\begin{equation}
    \mathrm{BR}(\mu\to e\gamma\gamma)\simeq \frac{\alpha^2}{640\pi^2}\frac{v_\text{ew}^6}{v_\Phi^2} \frac{m_\mu^2}{m_a^4 m_E^2}\,.
\end{equation}
Setting as the benchmark value the mass of the vector-like leptons to be $m_E = 10^4 \,v_\Phi$, we show in \cref{fig:alp_plots_combined} the contour in the $(m_a, v_\Phi^{\eminus1})$ plane, that corresponds to the branching ration $\mathrm{BR}(\mu\to e\gamma\gamma)= 10^{\eminus14}$ (dashed line).\footnote{Note that this constitutes a somewhat aggressive projection, given the present MEG results~\cite{MEG:2016leq, MEG:2020zxk}.} The off-shell regime, $m_a > m_\mu -m_e$, is to the right of the red vertical line.  
Despite the phase space and off-shellness penalties, for (sub)GeV ALPs, the expected experimental sensitivity reaches symmetry-breaking scales in the TeV range, and thus generically both $E$ and $\Psi$ can be above the limits from direct searches.\footnote{In our KSVZ-like UV completion for the lepton-flavor violating ALP~\cite{Kim:1979if,Shifman:1979if}, the  mass of the heavy fermion $\Psi$ is given by $m_\Psi = y_\Phi v_\Phi / \sqrt{2}$. Perturbativity together with direct 
search limits on $\Psi$ production~\cite{CMS:2024bni} thus impose a lower bound on the scale $v_\Phi$. 
In contrast, in DFSZ-like completions~\cite{Dine:1981rt, Zhitnitsky:1980tq, Calibbi:2020jvd} one can achieve, via tunings in the scalar potential,
 that the masses of new states are large enough, while  $v_\Phi$ remains small.}

\begin{figure}[t]
    \centering
	\begin{tikzpicture}[baseline=(a1.base), scale=0.35, transform shape]
			\begin{feynman}
			\vertex (a1) {\(\scalebox{2.4}{$\ell$}\)};
			\vertex [right = 3.2cm of a1](a2);
			\vertex [right = 2.6cm of a2](a3);
			\vertex [right = 2.6cm of a3](a4);
			\vertex [right = 3.2cm of a4](a5) {\(\scalebox{2.4}{$e$}\)};
			\vertex [above = 0.4cm of a3](a3below);
			\vertex [above = 3.6cm of a3](a3above);
			\vertex [above = 2.8cm of a3above](a3above1);
			\diagram*{
			(a1)--[fermion, thick](a2)--[fermion, thick](a4)--[fermion, thick](a5),
			(a2)--[dashed, thick, edge label=\(\scalebox{2.8}{$a\,$}\)](a3above)--[photon, thick, edge label=\(\scalebox{2.8}{$\,B$}\)](a4),
			(a3above)--[photon, thick, edge label=\(\scalebox{2.8}{$B\,$}\)](a3above1)
			};
			\draw[fill=black] (a2) circle(2.2mm);
			\draw[fill=black] (a4) circle(2.2mm);
			\draw[fill=black] (a3above) circle(2.2mm);
			\end{feynman}
	\end{tikzpicture}
    \caption{One-loop diagram illustrating the mixing of the ALP into the SMEFT $\cO_{eB}$ dipole operator. See \cref{sec:model_off_shell_ALP} for details.}
    \label{fig:ALP_SMEFT_RGE}
\end{figure}
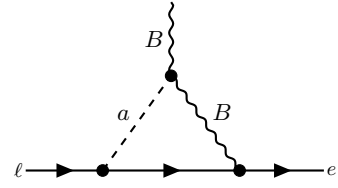

At one loop, the flavor-violating ALP--lepton operator in \cref{eq:ALP_mu_e} mixes into the flavor-violating SMEFT dipole operator~\cite{Galda:2021hbr} (see \cref{fig:ALP_SMEFT_RGE}). This generates the $\mu \to e\gamma$ decay with a branching ratio that, in the leading-log approximation, is given by\footnote{We do not include the subleading finite matching contributions~\cite{Davidson:2020ord, Fortuna:2022sxt, Fortuna:2023paj}.}
\begin{equation}
    \mathrm{BR}(\mu\to e\gamma)\simeq \frac{27\alpha^3}{256\pi^3}\frac{v_\text{ew}^6}{m_\mu^2}\frac{1}{m_E^2 v_\Phi^2}\log^2\lzm \frac{v_\Phi}{m_a} \dzm\,.
\end{equation}
In \cref{fig:alp_plots_combined}, we indicate the present experimental bound $\mathrm{BR}(\mu\to e\gamma)< 3.1 \times 10^{\eminus13}$~\cite{MEGII:2025gzr} by a dashed gray line, with the hatched region above denoting the excluded parameter space. This effect provides a stronger probe than $\mu \to e \gamma \gamma$, unless there is a tuned cancellation in $\mu \to e \gamma$ from another source. 

\begin{figure}[t]
    \centering
	\begin{tikzpicture}[baseline=(a1.base), scale=0.34, transform shape]
			\begin{feynman}
			\vertex (a1);
			\vertex [above left =3.4cm of a1] (a1ablef) {\(\scalebox{2.8}{$\ell$}\)};
			\vertex [below left =3.4cm of a1] (a1bellef) {\(\scalebox{2.8}{$H$}\)};
			\vertex [right =4.0cm of a1] (a2);
			\vertex [right =2.0cm of a1] (a1mid);
			\vertex [above =0.3cm of a1mid] (a1midAB) {\(\scalebox{2.8}{$E$}\)};
			\vertex [above right =3.4cm of a2] (a2abrig) {\(\scalebox{2.8}{$\ell$}\)};
			\vertex [below right =3.4cm of a2] (a2belrig) {\(\scalebox{2.8}{$H$}\)};
			\vertex [right =4.0cm of a2] (a3L);
			\vertex [right =4.0cm of a3L] (a3R);
			\vertex [right =4.0cm of a3R] (a4);
			\vertex [above right =3.4cm of a4] (a4AR) {\(\scalebox{2.8}{$\ell$}\)};
			\vertex [above left =3.4cm of a4] (a4AL) {\(\scalebox{2.8}{$\ell$}\)};
			\vertex [below left =3.4cm of a4] (a4BL) {\(\scalebox{2.8}{$H$}\)};
			\vertex [below right =3.4cm of a4] (a4BR) {\(\scalebox{2.8}{$H$}\)};
			\diagram*{
			(a1ablef)--[plain, thick](a1)--[dashed, thick](a1bellef),
			(a1)--[double, thick](a2),
			(a2)--[plain, thick](a2abrig),
			(a2)--[dashed, thick](a2belrig),
			(a4AR)--[plain, thick](a4)--[plain, thick](a4AL),
			(a4)--[dashed, thick](a4BL),
			(a4)--[dashed, thick](a4BR)
			};
			\draw[fill=black] (a1) circle(2.2mm);
			\draw[fill=black] (a2) circle(2.2mm);
			\draw[->, thick] (a3L) -- (a3R);
			\draw[fill=red!90!gray] (a4) +(-2.6mm,-2.6mm) rectangle +(2.6mm,2.6mm);
			\end{feynman}
	\end{tikzpicture}
    \caption{Matching of the heavy lepton $E$ onto the SMEFT Higgs–lepton current operators $\cO_{H\ell}^{(1)}$ and $\cO_{H\ell}^{(3)}$.}
    \label{fig:ALP_E_SMEFT}
\end{figure}
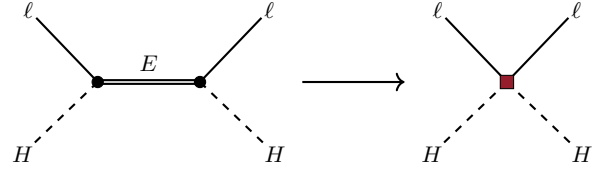

In addition, once $E$ is integrated out, it induces dimension-6 Higgs--fermion current operators at tree level (see~\cref{fig:ALP_E_SMEFT}), with flavor-violating Wilson coefficients of the form 
$[\cC_{H\ell}^{(1)}]_{12}=[\cC_{H\ell}^{(3)}]_{12}=-\theta_{e\mu}|y_{EH}|^2/4M_E^2$, where the corresponding SMEFT operators are, respectively, $[\cO_{H\ell}^{(1)}]_{12}=(H^\dagger i \overset{\text{\footnotesize$\leftrightarrow$}}{D}_\mu  H)(\bar \ell_1 \gamma^\mu  \ell_2)$ and  $[\cO_{H\ell}^{(3)}]_{12}=(H^\dagger i \overset{\text{\footnotesize$\leftrightarrow$}}{D^a_\mu}  H)(\bar \ell_1 \gamma^\mu \sigma^a  \ell_2)$, with $\sigma^a$ acting on $SU(2)_L$ indices~\cite{deBlas:2017xtg}. These operators feed into observables such as the coherent $\mu \to e$ conversion in nuclei and $\mu \to 3e$ decays, which impose important indirect constraints~\cite{Calibbi:2021pyh}. With $y_{EH}=1$, the current bounds translate into a limit $m_E \gtrsim 10^6\,\gev$. While stringent, current constraints on this model from $\mu\to e $ conversion are still less severe than the limits from $\mu\to e\gamma$, see also \cref{fig:alp_plots_combined}.

\vspace{0.2cm}
\noindent
{\textbf{On-shell ALP.}}
For $m_a< m_\mu$, the ALP is produced on-shell in a two-body $\mu \to e a$ decay. The experimental signature then depends on the ALP lifetime: for long lifetimes $a$ escapes the detector, resulting in a $\mu \to e +$inv signature, while if it decays via $a\to 2\gamma$, the signature is $\mu \to e 2\gamma$. Assuming that $a \to 2\gamma$ is the main decay channel, the effective $\mu\to e 2\gamma$ branching ratio measured by an experiment of typical size $L_\sscript{det}$ is given by
\begin{equation}
    \mathrm{BR}(\mu\to e\gamma\gamma)_\sscript{eff}=\mathrm{BR}(\mu\to e a) \times\Big[1 - \exp\big(\eminus {L_{\sscript{det}}}/{ \beta \gamma  c \tau_a}\big) \Big]\,,
\end{equation}
with $\beta$ the ALP velocity, $\gamma$ the corresponding Lorentz factor, and $\tau_a$ the ALP lifetime~\cite{Heeck:2017xmg} 
\beq
c\tau_a=74\,\mathrm{mm}\,\bigg(\frac{v_\Phi}{\gev}\bigg)^2\bigg(\frac{10\,\mev}{m_a}\bigg)^3\,.
\eeq 
In contrast, the $\mu \to e +$inv signature has the effective branching ratio
\begin{equation}
    \mathrm{BR}(\mu\to e+\mathrm{inv}) = \mathrm{BR}(\mu\to e a) \times \exp(\eminus L_{\sscript{det}}/\beta \gamma  c \tau_a)
    \,.
\end{equation}
In the limit $m_e\ll m_\mu$, the $\mu\to e a$ branching ratio is given by~\cite{Heeck:2017xmg} 
\begin{equation}
    \mathrm{BR}(\mu\to e a)\simeq\frac{3\pi^2}{2}\frac{v_\text{ew}^6}{m_\mu^4}\frac{y_{EH}^2 y_{E\Phi}^2}{m_E^2}\lzm1-\frac{m_a^2}{m_\mu^2}\dzm^2\,.
\end{equation}

In \cref{fig:alp_plots_combined} we show the experimental reach in the $(m_a, v_\Phi^{\eminus1})$ plane in the $m_a<m_\mu$ regime for two extreme cases: {\em(i)} when $a\to 2\gamma$ decay are within the detector volume and $\mathrm{BR}(\mu\to e\gamma\gamma)=10^{\eminus14}$ can be measured (dashed line), and {\em(ii)} when $a$ does not decay in the detector volume, assuming a rather optimistic experimental reach of $\mathrm{BR}(\mu\to e+\mathrm{inv})=10^{\eminus8}$ (solid line), for the $m_E=10^4 \,v_\Phi$ benchmark. Note that increasing $v_\Phi$ leads to longer ALP lifetimes, and, consequently, invisible decays dominate.\footnote{Whether or not the $\mu\to e +\mathrm{inv}$ decay gives the dominant sensitivity to on-shell ALP production, depends on the parameter values. For instance, for a benchmark with heavier vector-like leptons, e.g., $m_E = 10^{10}\,v_{\Phi}$, a distinct region at larger $m_a$ arises, in which $\mu \to e\gamma\gamma$ provides the leading sensitivity. However, in this case, the parameter space relevant for exotic muon decay branching ratios that can be observed in the next generation of experiments is already excluded by other searches.}

In summary, this benchmark highlights the complementarity between rare muon decays accessible at MEG~II and Mu3e, as well as constraints from astrophysical, beam-dump, and collider searches.

\section{Conclusions} 
\label{sec:conclusions}

\begin{figure}[t]
	\centering
	\includegraphics[width=0.46\textwidth]{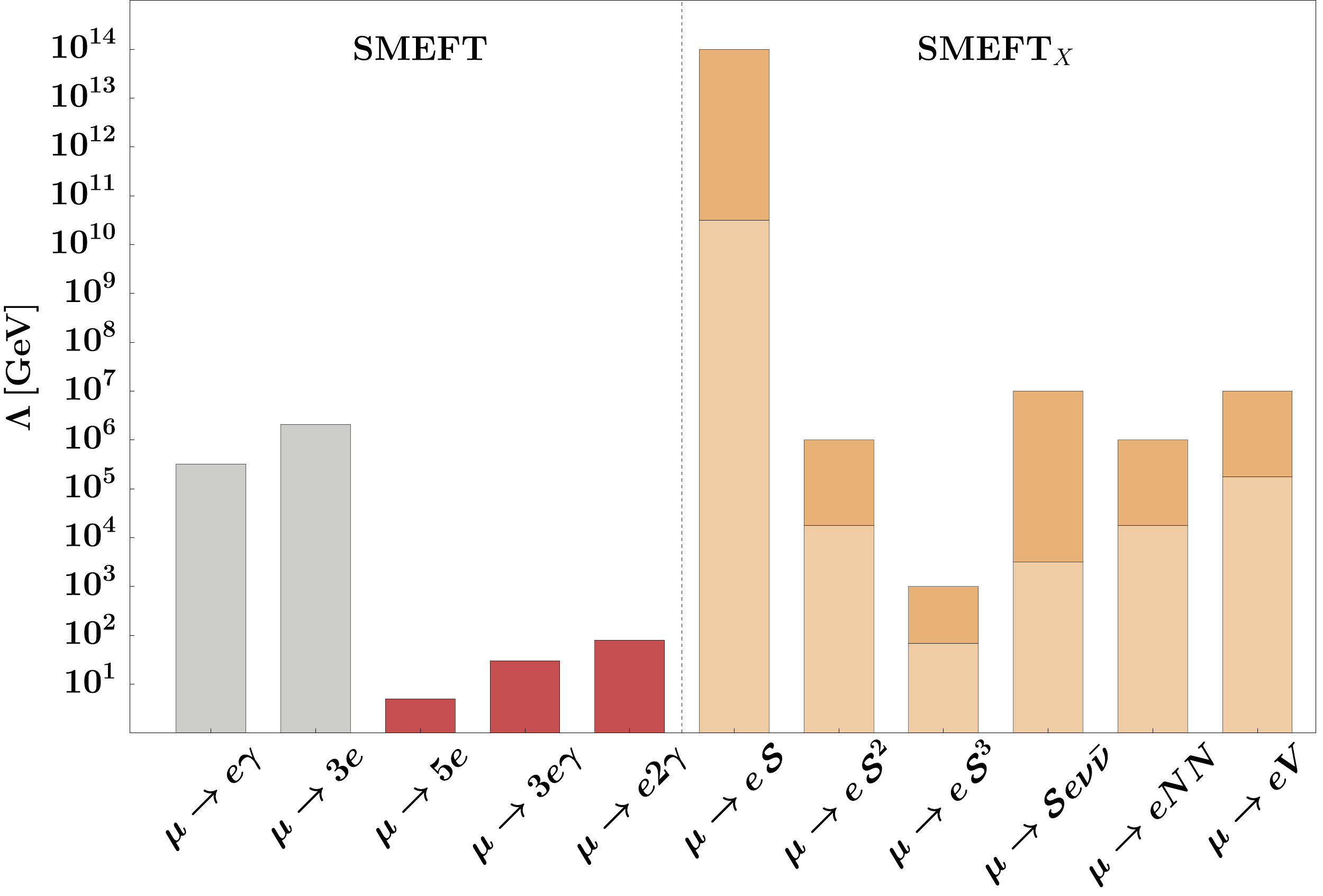}
    \caption{Summary of the new-physics reach in terms of effective scale $\Lambda$, for various muon decay channels induced by representative higher dimension operators, either assuming only heavy NP states that match onto SMEFT (gray and red bars, operators listed in \cref{tab:SMEFT_bounds}), or allowing for on-shell production of light new states, scalar singlet $\cS$, heavy neutral lepton $N$ or a vector $V$  (orange bars, SMEFT${}_X$ operators listed in \cref{tab:Leff:light}).  Dark (light) orange bars correspond to an assumed branching ratio of $10^{\eminus15}$ ($10^{\eminus8}$), illustrating the contrast between searches that are background-free (e.g. $\mu \to 5e$) and those that can become background-limited (e.g. $\mu \to e +\mathrm{inv}$), depending on the decay properties of $X=\cS, N, V$. For SMEFT-induced decays, we instead always assumed a branching ratio reach of $10^{\eminus15}$.}
    \label{fig:EFT_comparison_plot}
\end{figure}

\begin{table*}[t]
\centering
\renewcommand{\arraystretch}{1.2}
\scalebox{1.00}{
\begin{tabular}{lccccccccc}
\midrule
\midrule
\textbf{Model}
& ~~{\color{black}$\mu \to e \gamma$}~~
& ~~{\color{black}$e+$inv}~~
& ~{\color{black}$\mu-e$ conv.}~
& ~~~{\color{black}$3e$}~~~
& ~~{\color{black}$3e+$inv}~~
& ~~{\color{red}$e2\gamma$}~~
& ~~~{\color{red}$5e$}~~~
& ~~{\color{red}$5e+$inv}~~
& ~~~{\color{blue}$7e$}~~~ \\
\midrule
Flavor protect. scalar
&  & (\cmark) &  & (\cmark) &  &  & (\cmark) &  & \cmark \\
Dark photon + LFV
&  & (\cmark) &  & (\cmark) &  & & \cmark &  &  \\
Thermal inel. DM
&  & (\cmark) &  &  & \cmark  &  & (\cmark) & (\cmark) &  \\
On-shell ALP
& (\cmark) & \cmark & (\cmark) &  &  & (\cmark) &  &  &  \\
Off-shell ALP
& \cmark &  & (\cmark) &  &  & (\cmark) &  &  &  \\
Gauged $U(1)_{\mu-e}$
&  & \cmark &  & (\cmark) & \cmark &  & \cmark &  &  \\
\midrule
\midrule
\end{tabular}
}

\caption{Exotic muon decays with signatures composed of electrons, photons, and/or missing energy that can be expected in sample new physics models that have light new physics states below or close to the muon mass, see main text for further details. Checkmarks in parentheses denote that a signature is possible, but is subleading in the region of parameter space that this paper focuses on.}
\label{tab:summary_of_the_models_var_extended2}
\end{table*}

The ongoing and planned high-precision muon experiments at PSI, Fermilab, and JPARC are probing one of the core structural features of the SM, the conservation of charged lepton flavor. While most of the current experimental program focuses on the three golden channels, $\mu \to e \gamma$, $\mu \to 3e$, and $\mu\to e $ conversion, there are several other exotic muon decays, summarized in \cref{tab:exotic_signatures}, that can offer a complementary window to UV dynamics. 

As we showed in \cref{sec:EFT_considerations}, such multi-lepton/multi-photon signatures almost inevitably require the existence of light new physics, preferably produced on-shell. This finding is summarized in \cref{fig:EFT_comparison_plot}, which illustrates the sensitivity to the effective scales for selected SMEFT (left) or SMEFT${}_X$ (right)
higher-dimension operators, for different decay topologies. In the construction of the SMEFT Lagrangian, all new physics states are assumed to be heavy and can thus be integrated out. For SMEFT${}_X$, one instead assumes that there exists a light new physics state $X$ that can be produced on-shell in muon decays. In the SMEFT limit, only the golden cLFV muon decay channels probe a high effective NP scale $\Lambda$, well above the EWSB scale. The definition of the effective scale $\Lambda$ for each of the considered SMEFT operators, including the size of the dimensionless Wilson coefficient, is given in \cref{tab:SMEFT_bounds}. 

The situation is qualitatively different for SMEFT${}_X$; now, also the high-multiplicity decay modes can be sensitive to very high new physics scales. The achievable reach on $\Lambda$ depends on the decay properties of the light NP state $X$. For example, for the searches that are effectively background-free, such as the $\mu \to 5e$ signature, that is a result of a decay chain $\mu \to e 2X$ and $X\to 2e$, we illustrate the reach in \cref{fig:EFT_comparison_plot} by assuming that a sensitivity to a branching ratio of $10^{\eminus15}$ can be reached (dark orange bars). For the decay channels that are expected to be background-limited, such as $\mu \to e + \mathrm{inv}$, from $\mu \to e X$, with $X$ escaping detection, we instead take as an illustrative benchmark that a branching ratio of  $10^{\eminus8}$ can be reached (light orange bars). We observe that even within this simplistic assumption, there are significant differences between the different assumed decay channels in the size of the NP scale $\Lambda$ being probed. In general, the higher the dimension of the SMEFT${}_X$ operator that generates the exotic decay, the lower the scale $\Lambda$ that is being probed for the same exotic branching ratio value. This simplified analysis should, of course, be used with caution; its primary purpose is merely to point out that several high-multiplicity channels merit further exploration, both by performing more detailed studies of the expected experimental sensitivities at various rare muon decay experiments, as well as by conducting the actual experimental searches.

The experimental search strategies may gain further sensitivity if one can utilize kinematical constraints, such as the exotic muon decays being due to a particular decay chain topology. For this purpose, we construct four benchmark models in \cref{sec:models}. For such explicit models, one can also evaluate the importance of other constraints, such as existing collider searches, beam dump experiments, astrophysical constraints, and constraints from other rare decays.  The four benchmark models, the flavor protected scalar, Higgsed dark photon model with higher dimension source of LFV, thermal inelastic DM model (that also contains a source of higher-dimension LFV operator), and on-shell and off-shell ALPs, as well a variation of the dark photon benchmark model involving $U(1)_{\mu-e}$ gauge group (with additional details provided in \cref{sec:AppU(1)mu-e}) lead to a large variety of different exotic signatures, as summarized in \cref{tab:summary_of_the_models_var_extended2}. For concreteness, we also provide the corresponding \texttt{UFO} files for the scalar benchmark model in a {\tt github} repository~\cite{ExoticMuonsGit}.

A discovery of one of the above exotic decays could give valuable insights into the dynamics at very short distances. 
For instance, in the first benchmark models, flavor symmetries ensure that $\mu \to 7e$ decay is the leading exotic decay channel, while lower-multiplicity modes are suppressed. If such a pattern were to be realized experimentally, it would point to a specific mechanism of flavor symmetry breaking. The third benchmark model, instead, is connected to a low-mass thermal dark matter setup that could, surprisingly, be first discovered in high-multiplicity exotic muon decays. The second benchmark model is connected to the third one, but does not contain the dark matter sector, so that the phenomenology is limited to that of a dark photon accompanied by flavor-violating interactions due to higher-dimensional operators. A variant of this model based on $U(1)_{\mu-e}$ gauge group is given in Appendix \ref{sec:AppU(1)mu-e}.

In conclusion, although we have presented only a few representative models, the study we performed demonstrates that searches for high-multiplicity rare muon decays can be interesting probes of feebly coupled sectors.

\vspace{0.2cm}
\noindent
{\textbf{Acknowledgments.}}
We thank Xavier Ponce D\'iaz, Anders Eller Thomsen, and Alessandro Valenti for useful discussions. This work has received funding from the Swiss National Science Foundation (SNF) through the Eccellenza Professorial Fellowship ``Flavor Physics at the High Energy Frontier,'' project number 186866. JZ acknowledges support in part by DOE grant DE-SC101977,  DE-SC0026301, and by NSF grants OAC-2103889, OAC-2411215, and OAC-2417682.

\appendix 
\renewcommand{\thesection}{\Alph{section}}
\renewcommand{\thesubsection}{\Alph{section}.\arabic{subsection}}
\setcounter{section}{0}


\section{Dark sectors gauged under $U(1)_{\mu-e}$}
\label{sec:AppU(1)mu-e}

In this appendix, we explore the possibility that the SM leptons as well as the dark sector fields are gauged under $U(1)_d$, extending the discussion in \cref{sec:DarkPhoton,sec:model_dark_higgs}. For exotic muon decays, perhaps the most phenomenologically compelling case arises when the new gauge group is $U(1)_{\mu-e}$, which we therefore take as our focus.

\subsection{Gauged $U(1)_{\mu-e}$ model}
Under $U(1)_{\mu-e}$, the SM leptons carry the vector-like charges $[\ell_1]=[e_1]=-[\ell_2]=-[e_2]=-1$ and $[\ell_3]=[e_3]=0$. Similarly to the dark photon model in \cref{sec:DarkPhoton}, we also extend the field content by an additional scalar $\Phi$, which is taken to have a $U(1)_{\mu-e}$ charge of $[\Phi]=+2/n$, where $n$ is an integer (a choice to be made clear below). The  $U(1)_{\mu-e}$ is spontaneously broken when $\Phi$ obtains a VEV $v_\Phi$, with $\Phi=(v_\Phi+\varphi)/\sqrt 2$, and $\varphi$ a light dark Higgs. The mass of the $U(1)_{\mu-e}$ gauge boson $V_\mu$ is then $m_V=g_V v_\Phi$, where $g_V$ is the gauge coupling, which is experimentally bounded to be small ($g_V\ll 1$). In principle, there is also kinetic mixing between $V_\mu$ and SM hypercharge, as in the dark photon model in \cref{sec:DarkPhoton}. Typically, the kinetic mixing term leads to subleading effects compared to the direct $U(1)_{\mu-e}$ couplings to leptons, and we therefore neglect it in the remainder of this section.

As far as the muon decays are concerned, the $U(1)_{\mu-e}$  model has similar phenomenology as the dark photon model in \cref{sec:DarkPhoton}, however, the two models do differ in details. Most importantly, in the dark photon model, the gauge boson couples more strongly to dark sector particles than to visible particles, i.e., $g_V\gg \epsilon g_V$, while in the $U(1)_{\mu-e}$ model, $V_\mu$ couples to both sectors with similar couplings ($\sim g_V$). This means that $g_V$ is bounded in the $U(1)_{\mu-e}$ model to be below roughly $g_V\lesssim 10^{\eminus4}$. 

In both dark photon model and the $U(1)_{\mu-e}$ model, we are interested in the regime $2 m_e<m_V<(m_\mu-m_e)/2$. Since $g_V\ll1$ in $U(1)_{\mu-e}$  model, this implies that $v_\Phi$ is now well above $m_\mu$, while in dark photon model $g_V$ can be sizable and thus $v_\Phi$ could be in the GeV regime. In both models, the dark Higgs mass is given by $m_\varphi=\sqrt{2 \lambda_\Phi} v_\Phi$, where $\lambda_\Phi$ is the quartic in the dark Higgs potential, defined by $V_\Phi=\lambda_\Phi (\Phi^\dagger \Phi-v_\Phi^2/2)^2$. Since $v_\Phi\gg m_\mu$, we require a small quartic ($\lambda_\Phi\ll 1$), so that the dark Higgs is light ($m_\varphi\ll v_\Phi$), where, in particular, we are interested in the regime when $\varphi$ is lighter than the muon. Instead, in the dark photon model in \cref{sec:DarkPhoton}, it was possible to achieve this mass hierarchy also for sizable values of $\lambda_\Phi$, but correspondingly smaller values of $v_\Phi$.

The other difference with respect to the dark photon model is the form of the flavor-violating higher-dimension operators,
\begin{equation}
\begin{split}
    -\cL\supset 
   &\sum_{i=e,\mu} \hat y_{ii}\,\bar\ell_i He_i
    +y_5^{e\mu}\biggr(\frac{\Phi^\dag}{\Lambda}\biggr)^n\big(\bar\ell_1He_2\big)
    \\
    &+y_5^{\mu e}\biggr(\frac{\Phi}{\Lambda} \biggr)^n \big(\bar\ell_2 H e_1\big)
    +\hermc\,.
    \end{split}
    \label{eq:Umu-e:Lagr}
\end{equation}
Note that, unlike in the $U(1)_d$ case, which required two insertions of $\Phi$ field in the cLFV operators, see \cref{eq:cLFV:portal}, the $U(1)_{\mu-e}$ model allows also for a single insertion of $\Phi$ field (depending on the quantum numbers; linear $\Phi$ insertion occurs for $n=1$). In addition, note that since the $U(1)_{\mu-e}$ charge assignments are vector-like, the renormalizable flavor-diagonal SM Yukawa interactions for charged leptons are allowed, the same as in the dark photon model. 

The other difference with respect to the $U(1)_d$ model is that the gauge boson $V$ now decays via $V\to 2e$ as well as via $V\to 2\nu_e$, with the branching ratios satisfying $\mathrm{BR}(V\to 2e): \mathrm{BR}(V\to 2 \nu_e)=2:1$. That is, roughly 33\% of $V$ decays are into invisible final state, while for $U(1)_d$ the  $V\to 2e$ decay channel is almost 100\%. This feature can be used to distinguish between the two models experimentally. 

Once the Higgs and the scalar $\Phi$ acquire VEVs, the electroweak symmetry and the $U(1)_{\mu-e}$ gauge symmetry are spontaneously broken, respectively. Setting the Higgs and the scalar $\Phi$ in the dimension-5 operators to their VEVs generates off-diagonal entries in the charged lepton mass matrix.  In the small mixing limit, the charged lepton mass matrix is diagonalized by rotations of the left- and right-handed fields, where
\begin{equation}
	\theta_{L} \simeq \frac{y_5^{e\mu}}{y_{\mu}}\biggr(\frac{v_\Phi}{\sqrt{2}\Lambda}\biggr)^n\,,
    \qquad 
    \theta_{R} \simeq \frac{y_5^{\mu e}}{y_{\mu}}\biggr(\frac{v_\Phi}{\sqrt{2}\Lambda}\biggr)^n\,.
\end{equation}
The misalignment between the $U(1)_{\mu-e}$ interaction and the mass basis implies that $V_\mu$ has flavor-violating couplings to charged leptons:
\beq
	\cL_{V}\supset 2 g_V\theta_{L} (\bar e_L  \slashed V \mu_L) + 2 g_V\theta_{R} (\bar e_R  \slashed V \mu_R)+\hermc\,.
\eeq
In the $m_e, m_V\ll m_\mu$ limit, the branching ratio for $\mu\to eV$ decays is given by
\beq
{\rm BR}(\mu \to e V)=\frac{1}{4\pi}\frac{m_\mu}{\Gamma_\mu} g_V^2 \big(\theta_L^2+\theta_R^2\big)\,.
\eeq
The Lagrangian in \cref{eq:Umu-e:Lagr} also leads to flavor-violating couplings of the light dark Higgs:
\beq
-\cL\supset n\varphi \Big[ \theta_L \frac{m_\mu}{v_\Phi} \big(\bar e P_R \mu\big)+\theta_R \frac{m_\mu}{v_\Phi} \big(\bar e P_L \mu\big)\Big] +\hermc\,.
\eeq
If the dark Higgs is lighter than $m_\mu-m_e$, the above interaction induces a $\mu\to e \varphi$ decay. In the $m_\varphi\ll m_\mu$ limit we have
\beq
\begin{split}
\label{eq:Br:muphi:Umue}
{\rm BR}(\mu \to e \varphi)= \frac{1}{16\pi }\frac{m_\mu}{\Gamma_\mu}\Big(n\frac{m_\mu}{v_\Phi}\Big)^2\big(\theta_L^2+\theta_R^2)\,.
\end{split}
\eeq
Using the fact that the $V_\mu$ mass is $m_V=g_V v_\Phi$, the two branching ratios satisfy a simple relation:
\beq
\begin{split}
{\rm BR}(\mu \to e \varphi)= \frac{1}{4}\Big(\frac{n m_\mu}{m_V}\Big)^2{\rm BR}(\mu \to e V)\,.
\end{split}
\eeq
Moreover, if $m_V\ll m_\mu/2$, and the $\mu\to e \varphi$ decays are kinematically allowed, then the $\mu\to e \varphi$ will dominate over $\mu\to e V$. This then leads to $\mu\to 5e$ signature via the decay chain $\mu\to e\varphi$, with $\varphi\to 2V$ and $V\to 2e$, cf. \cref{fig:Dark_photon_model}. If, on the other hand, $\varphi$ is too heavy for $\mu\to e \varphi$ decay to be kinematically allowed, then $\mu\to e V$ is the dominant exotic decay, leading to the $\mu\to 3e$ signature.

Since for $n=1$ the coupling to $\Phi$ in LFV interaction \cref{eq:Umu-e:Lagr} is linear, the $\mu\to 5e$ searches will probe much higher scales than in the dark photon model (these are shown in \cref{fig:DM_plot}). Taking $y_5^{e\mu},y_5^{\mu e}\sim y_\mu$ and $m_\mu\gg m_e$, the branching ratio for $\mu \to e \varphi$ decay, \cref{eq:Br:muphi:Umue}, becomes
\beq
\begin{split}
{\rm BR}(\mu \to e \varphi)\sim \frac{1}{16\pi }\frac{1}{\Gamma_\mu}\frac{m_\mu^3}{ \Lambda^2}\,,
\end{split}
\eeq
which for ${\rm BR}(\mu \to e \varphi)\lesssim 10^{\eminus15}$ then translates to $\Lambda\gtrsim 3\times 10^{14}\,\gev$.

\subsection{Inelastic thermal DM and  $U(1)_{\mu-e}$}
In principle, the inelastic thermal DM model in \cref{sec:model_dark_higgs} can be extended to the $U(1)_{\mu-e}$ case as well. However, while in the inelastic DM model with dark photon, the relic abundance was set by the $\chi_1\chi_2\to V^*\to e^+e^-$ coannihilation cross section \cite{Duerr:2019dmv}
\beq
\sigma v(\chi_1\chi_2\to e^+e^-)\approx \frac{4\pi \epsilon^2 \alpha_\sscript{em} \alpha_V m_{12}^2}{(m_{12}^2-m_V^2)^2+m_V^2\Gamma_V^2}\,,
\eeq
where $m_{12}=m_{\chi_1}+m_{\chi_2}$ and $\alpha_V=g_V^2/4\pi$, in the $U(1)_{\mu-e}$ model the corresponding cross section is given by
\beq
\sigma v(\chi_1\chi_2\to e^+e^-)\approx \frac{4\pi \alpha_V^2 m_{12}^2}{(m_{12}^2-m_V^2)^2+m_V^2\Gamma_V^2}\,.
\eeq
Phenomenologically, this constitutes a significant change: while in the $U(1)_d$ model one has two parameters that control the strength of interactions with the SM, namely $\varepsilon$ and $g_V$, and can thus take $\varepsilon\ll g_V\sim {\mathcal O}(1)$, in case of $U(1)_{\mu-e}$ the laboratory searches require $g_V\ll 1$. Obtaining the correct relic abundance would thus require significant tuning, with $m_{12}$ very close to $m_V$, typically $(m_{12}-m_V)/m_V\lesssim 10^{\eminus8}$. We thus do not consider the inelastic thermal DM with the $U(1)_{\mu-e}$ gauge boson mediator as a viable benchmark model.

\bibliographystyle{JHEP}
\bibliography{refs.bib}

\end{document}